\documentclass[aps,prl,reprint,superscriptaddress]{revtex4-1}
\usepackage{amsmath}
\usepackage{amssymb}
\usepackage{graphicx}
\usepackage{dcolumn}
\usepackage{bm}
\usepackage{xcolor}
\usepackage{graphicx}
\usepackage{tikz-cd}
\usepackage{natbib}
\usepackage{siunitx}

\begin{document}
\preprint{APS/123-QED}
\title{Second Quantization Approach to Many-Body Dispersion Interactions:\\  Implications for Chemical and Biological Systems}
\author{Matteo Gori}
\email[Corresponding email address: ]{matteo.gori@uni.lu}
\affiliation{Department of Physics and Materials Science, University of Luxembourg, L-1511 Luxembourg City, Luxembourg}
\affiliation{Quantum Biology Laboratory, Howard  University, Washington DC 20060, USA} \author{Philip Kurian}
\email[Corresponding email address: ]{pkurian@howard.edu}
\homepage[Website: ]{https://quantumbiolab.com}
\affiliation{Quantum Biology Laboratory, Howard  University, Washington DC 20060, USA}
\author{Alexandre Tkatchenko}
\email[Corresponding email address: ]{alexandre.tkatchenko@uni.lu}
\affiliation{Department of Physics and Materials Science, University of Luxembourg, L-1511 Luxembourg City, Luxembourg}

\begin{abstract}
The many-body dispersion (MBD) framework
is a successful approach for modeling the long-range electronic correlation energy and optical response of systems with thousands of atoms. Inspired by field theory, here we develop a second-quantized MBD formalism (SQ-MBD) that recasts a system of atomic quantum Drude oscillators in a Fock-space representation. SQ-MBD provides: (\emph{i}) tools for projecting observables (interaction energy, transition multipoles, polarizability tensors) on coarse-grained representations of the atomistic system ranging from single atoms to large structural motifs, (\emph{ii}) a quantum-information framework to analyze correlations and (non)separability among fragments in a given molecular complex, and (\emph{iii}) a path toward the applicability of the MBD framework to molecular complexes with millions of atoms. The SQ-MBD approach offers novel insights into quantum fluctuations in molecular systems and enables direct coupling of collective plasmon-like MBD degrees of freedom with arbitrary environments, providing a tractable computational framework to treat dispersion interactions and polarization response in intricate systems.
\end{abstract}
\maketitle
Noncovalent interactions \cite{stone2013theory,hirschfelder2009intermolecular,margenau2013theory,kaplan2006intermolecular} 
play a key role in determining 
physicochemical properties, given that they influence the structure~\cite{hoja2019reliable}, 
stability~\cite{mortazavi2018structure,hoja2018first}, dynamics~\cite{reilly2014role,stohr2019quantum,galante2021anisotropic}, and electric~\cite{kleshchonok2018tailoring} and optical~\cite{ambrosetti2022optical} responses in a wide range of molecules and materials~\cite{vdWDF-review,Grimme-CR,Hermann-CR}.
In particular, van der Waals (vdW) 
dispersion interactions and long-range electron 
correlation energy must 
be treated with quantitative many-body 
methods~\cite{parsegian2005van,Dobson-Gould,Kresse-RPA,Xinguo-RPA,tkatchenko2015current,woods2016materials,mahanty1973dispersion,richardson1975dispersion,paranjape1979van}.
Different methods \cite{carlsson1997exchange,hennig2001density,schade2017reduced,piris2017global,piris2021global} have been proposed to include dispersion 
interactions in the form of non-local vdW density 
functionals.
The many-body dispersion (MBD) framework~\cite{tkatchenko2012accurate,ambrosetti2014long}
has been firmly established as an efficient and 
accurate approach.
In MBD, the electronic response 
properties of each atom are represented by a quantum Drude oscillator (QDO)~\cite{jones2013quantum}.
The long-range correlations among the electronic fluctuations emerge from 
the dipolar coupling between the QDOs.
The MBD method can be now routinely applied to systems with up to
$N\sim 10^4$ atoms~\cite{stohr2019quantum}, a size limitation owing to the $N^3$ computational scaling of MBD. 
Furthermore, MBD effects have been shown to extrapolate to mesoscale processes~\cite{Prashanth-PRL,Prashanth-SciAdv}, including solvation and folding of proteins~\cite{stohr2019quantum,Piquemal-MBD1} and the delamination of graphene from surfaces~\cite{hauseux2020quantum}, demonstrating the interplay between MBD modes and collective nuclear vibrations~\cite{Prashanth-SciAdv,hauseux2022colossal}.
These findings suggest that MBD interactions contribute to cooperative effects between electronic and nuclear degrees of freedom in complex chemical and biophysical systems. These effects include non-local allosteric pathways in enzymes from coordinated electronic fluctuations~\cite{pingoud2014type,kurian2016quantum, kurian2018water} 
and the emergence of giant electric-dipole oscillations in biomolecules that mediate long-range intermolecular interactions~\cite{nardecchia2018out,lechelon2022}. 
Pursuing the study of MBD effects in realistic systems in complex environments requires simulations with 
millions of atoms, which are infeasible at the moment even with stochastic implementations~\cite{Piquemal-MBD1,Piquemal-MBD2}. 
The development of a coarse-grained MBD model would be
a compelling strategy to provide a conceptual and computational leap to extend the applicability of MBD to million-atom systems. 
With this goal in mind, we propose here a second quantization formulation of the MBD model (SQ-MBD) that
considerably simplifies the calculation 
of the fragment contributions to observables 
stemming from collective MBD modes, and enhances physical intuition on how MBD effects operate 
to connect different length scales in macroscopic systems.
For instance, the fragment contribution to
the total MBD energy could be used as reference data
for machine-learned force fields~\cite{chmiela2019sgdml,noe2020machine,musil2021physics}, while coarse-grained fragment contributions to transition dipole elements of excited MBD states could yield effective models to predict collective optical response in biomolecular complexes~\cite{pearlstein1991theoretical,shi2017generation, celardo2019existence} and J-aggregates~\cite{ma2021organic, eisele2012}.

We consider a system of $N$ atoms with fixed nuclear positions 
$\{\boldsymbol{R}_A\}_{A=1}^{N}$. 
The valence electronic response properties of atom $A$
are described by a 3D isotropic QDO parametrized by 
its angular frequency $\omega_A$, mass $m_A$, and
electric charge $Z_A e$. 
The MBD Hamiltonian 
\begin{equation}
\label{eq:HMBD_IQ_pq}
    \hat{H}_{\rm MBD} = \dfrac{1}{2}\sum_{A=1}^{N}\left[ \|\hat{\boldsymbol{p}}_A\|^2  + \sum_{B \neq A} \hat{\boldsymbol{q}}_A \mathbb{V}_{AB} \hat{\boldsymbol{q}}_B\right]
\end{equation}
describes the interacting system of QDOs, where 
$\hat{\boldsymbol{q}}_A=\sqrt{m_A}\hat{\boldsymbol{r}}_A=(\hat{q}_{A_{x_1}},\hat{q}_{A_{x_2}},\hat{q}_{A_{x_3}})$ 
is the (mass-weighted) displacement operator of the QDOs associated with atom $A$, and the 
momentum operator $\hat{\boldsymbol{p}}_A$ 
is the associated canonical conjugate variable.
\begin{figure*}[htb!]
        \centering
    \includegraphics[scale=0.095]{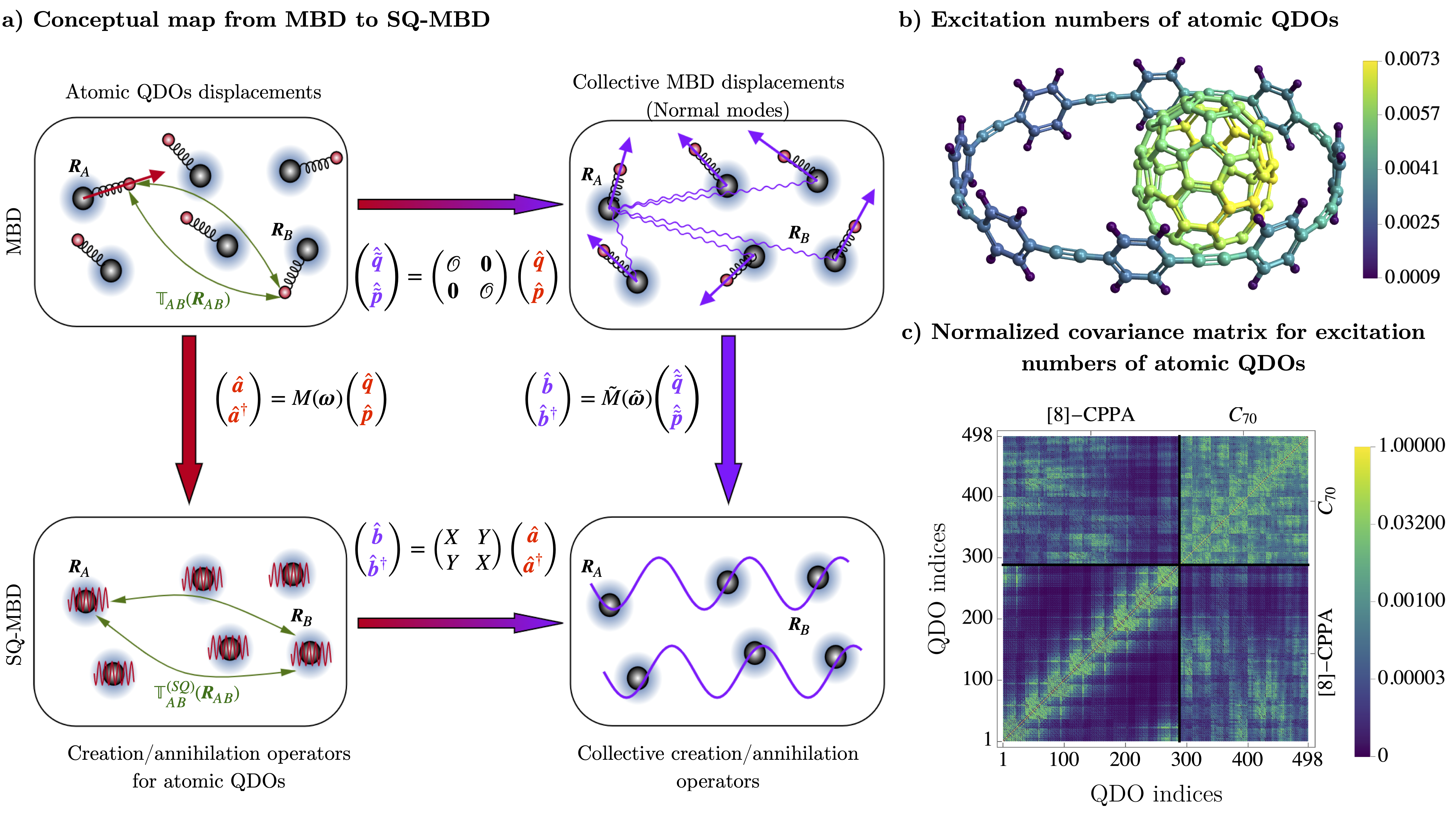}
  \caption{\textbf{Theory and practice of the SQ-MBD method}. 
  Panel (a) shows a schematic
  representation of the commutative diagram
  establishing the mapping between the original MBD framework and the second-quantized formalism (SQ-MBD). $\bm{M}$ and $\tilde{\bm{M}}$ denote transformation matrices between first- and second-quantization representations. 
Panel (b) shows the mean excitation numbers of atomic QDOs in the MBD ground state for the supramolecular complex of $C_{70}$ fullerene surrounded by a cycloparaphenyl ring composed of 8 units ([8]-CPPA). Panel (c) shows the normalized covariance matrix of excitation numbers for interacting atomic QDOs, decomposed into single Cartesian components. The index $3(A-1)+i$ is assigned to the QDO associated to the displacement of the Drude particle on the $A$-th atom along the $i$-th Cartesian direction.}
  \label{fig:ItoIIquantization}
\end{figure*}
The interaction potential is described by the $3\times 3$ matrices $\mathbb{V}_{AB}=\omega_A\omega_{B}\left[\mathbb{I}\delta_{AB}+\sqrt{\mathcal{A}_{A}^{(0)}\mathcal{A}_{B}^{(0)}}\,\,\mathbb{T}_{AB}(\boldsymbol{R}_{AB})\right]$ 
where $\mathbb{I}$ and $\mathbb{T}_{AB}$ are, respectively, the $3\times 3$
identity matrix and the dipole-dipole coupling between QDOs of atom $A$ and atom $B$ ($\boldsymbol{R}_{AB}=\boldsymbol{R}_B - \boldsymbol{R}_A$).
The Hamiltonian in Eq. \eqref{eq:HMBD_IQ_pq} is quadratic in the QDO variables, so it can be reduced to the normal form $
\label{eq:HMBD_IQ_mbd}
    \hat{H}_{\rm MBD} = (1/2)\left[\sum_{k=1}^{3N} \hat{\tilde{p}}_k^2  +\tilde{\omega}_k^2 \hat{\tilde{q}}_k^2 \right]
$
where $\hat{\tilde{q}}_k,\hat{\tilde{p}}_k,\tilde{\omega}_{k}$ are, 
respectively, the displacement, the momentum, and the angular frequency of
the $k$-th normal mode.  We assume in what
follows that $0<\tilde{\omega}_{k}\leq\tilde{\omega}_{k+1}$ for all $k=1,...,3N$. The canonical transformation from the 
atomic-based operators $\{\hat{\boldsymbol{q}}_A,\hat{\boldsymbol{p}}_A\}_{A}$ to the $3N$ MBD
normal-mode variables
$\{\hat{\tilde{q}}_k,\hat{\tilde{p}}_k\}_{a}$ is determined by the orthogonal matrix $\mathcal{O}$ that diagonalizes the potential matrix
in the MBD Hamiltonian (see Figure \ref{fig:ItoIIquantization}a).
In this framework, the MBD energy is given by the energy difference between the interacting QDOs and the QDOs at infinite separation, $ E_{\rm MBD}=(\hbar/2)\left[\sum_{k=1}^{3N}\tilde{\omega}_k -3\sum_{A=1}^{N} \omega_A\right]$.
Although the orthogonal matrix $\mathcal{O}$ and the set of eigenenergies $\hbar \tilde{\omega}_k$ fully characterize the MBD ground state, such a first quantization approach is not optimal. For instance, it is not straightforward to represent the ground and the excited collective eigenstates in the eigenbasis of the atomic QDOs. The calculation of expectation values of localized observables in the MBD ground state is also cumbersome. 

The second quantization framework (SQ-MBD) developed here overcomes these issues, providing a description of the degrees of freedom of the atomic QDOs and of the MBD collective plasmonic modes in terms of the algebra of ladder operators for isolated QDOs 
$\{\hat{a}_{A_{x_i}},\hat{a}^{\dagger}_{A_{x_i}}\}_{A,i}$
with the associated basis set 
$|\boldsymbol{n}\rangle=\bigotimes_{A,i} 
|n_{A_{x_i}}\rangle$, and the algebra of ladder 
operators for coupled QDOs $\{\hat{b}_{k},\hat{b}^{\dagger}_{k}\}_{k}$
with the basis set 
$|\tilde{\boldsymbol{n}}\rangle=\bigotimes_{k} 
|\tilde{n}_{k}\rangle$.
Owing to the linear transformation 
$M(\boldsymbol{\omega})$($\tilde{M}(\tilde{\boldsymbol{\omega}})$)
from atomic QDO (normal mode) displacements/momenta operators to the 
corresponding set of ladder operators for the 
collective MBD modes, it is possible to construct the commutative diagram reported in Figure 
\ref{fig:ItoIIquantization}a, with an 
explicit expression for the mapping between 
the two algebras of creation/annihilation operators, in terms of
the orthogonal matrix $\mathcal{O}$ and two sets of
eigenfrequencies $\{\omega_{A}\}_{A}$ and
$\{\tilde{\omega}_{k}\}_{k}$, given by
\begin{equation}
\label{eqref:Bogotrans}
    \begin{pmatrix}
    \hat{\boldsymbol{b}}\\
    \hat{\boldsymbol{b}}^{\dagger}\\
    \end{pmatrix}
    =
    \begin{pmatrix}
    X & Y\\ 
    Y & X\\
    \end{pmatrix}
    \begin{pmatrix}
    \hat{\boldsymbol{a}}\\
    \hat{\boldsymbol{a}}^{\dagger}\\
    \end{pmatrix}\,,
\end{equation}
where $X(\mathcal{O},\boldsymbol{\omega},\tilde{\boldsymbol{\omega}}),Y(\mathcal{O},\boldsymbol{\omega},\tilde{\boldsymbol{\omega}})$ are $3N \times 3N$ real matrices (see Supplemental Material for further details). The linear map in Eq.~\eqref{eqref:Bogotrans} is a multimodal Bogoliubov transformation \cite{berazin2012method,ripka1986quantum}, preserving the canonical commutation relations of the ladder operator algebra. Bogoliubov transformations in finite quantum systems admit a unitary representation \cite{ripka1986quantum} $\hat{S}^{-1}=\hat{S}^{\dagger}$ such that $\hat{b}_k^{(\dagger)}=\hat{S}\hat{a}^{(\dagger)}_{A_{x_i}}\hat{S}^{-1}$ for $k=3(A-1)+i$, connecting the ground state $|\boldsymbol{0}\rangle=\bigotimes_{A,i} 
|0_{A_{x_i}}\rangle$ of the uncoupled atomic QDO system with the collective MBD ground state
\begin{equation}
\label{eqref:Gs_QDOtoMBD}
    |\tilde{\mathbf{0}}\rangle=\hat{S}|\boldsymbol{0}\rangle=\dfrac{e^{\frac{1}{2} \sum\limits_{A,B=1}^N\sum\limits_{i,j=1}^3\hat{a}_{A_{x_i}}^{\dagger} \Theta_{A_{x_i} B_{x_j}} \hat{a}_{B_{x_j}}^{\dagger}}}{\mathrm{det}^{1/4}(XX^{T})} |\mathbf{0}\rangle\,,
\end{equation}
where $\Theta = X^{-1}Y $ is a $3N \times 3N$ symmetric real matrix. 
Equations \eqref{eqref:Bogotrans} and \eqref{eqref:Gs_QDOtoMBD} represent 
the information encoded in the MBD ground state in terms of the excited states of the atomic QDOs. The SQ-MBD Hamiltonian thus reads
\begin{equation}
    \hat{H}_{\rm SQ-MBD}=\sum_{k=1}^{3N} \hbar \tilde{\omega}_k \left(\hat{b}_{k}^{\dagger} \hat{b}_{k} +\dfrac{1}{2}\right)\,\,.
\end{equation}

In what follows, we present applications of the SQ-MBD framework to the analysis of the MBD ground state properties in supramolecular and biological systems. 
For all the results that we present, the atomic QDOs frequencies
$\boldsymbol{\omega}$, the MBD eigenfrequencies $\tilde{\boldsymbol{\omega}}$
and the orthogonal matrix of MBD eigenvectors $\mathcal{O}$ have all been derived
using the current state-of-the-art MBD@rsSCS approach (see Supplemental Material for 
further details). This ensures the consistency of our total SQ-MBD energies with the MBD@rsSCS method.

We first analyze the excitation numbers of atomic QDOs in the many-body state defined as $\langle \hat{\mathcal{N}}_{A}\rangle_{\tilde{\boldsymbol{0}}}=\sum_{i=1}^3 \langle \tilde{\boldsymbol{0}}| \hat{a}_{A_{x_i}}^{\dagger} \hat{a}_{A_{x_i}} |\tilde{\boldsymbol{0}} \rangle$
as well as the pairwise correlations between excitation numbers of atomic QDOs
$\mathrm{Cov}_{\tilde{\boldsymbol{0}}}(\hat{\mathcal{N}}_{A_{x_i}} \hat{\mathcal{N}}_{B_{x_j}})=\langle \hat{\mathcal{N}}_{A_{x_i}} \hat{\mathcal{N}}_{B_{x_j}} \rangle_{\tilde{\boldsymbol{0}}}-\langle \hat{\mathcal{N}}_{A_{x_i}}\rangle_{\tilde{\boldsymbol{0}}} \langle \hat{\mathcal{N}}_{B_{x_j}} \rangle_{\tilde{\boldsymbol{0}}}$, which are reported in Figure \ref{fig:ItoIIquantization}b,c for a complex of $\text{C}_{70}$ fullerene surrounded by an [8]-CPPA molecular ring. The $\text{C}_{70}$-CPPA system  constitutes a benchmark for the calculation of the dispersion energy \cite{stohr2021coulomb},
given that it is essentially homonuclear and highly polarizable. 
The correlations have been normalized as follows  $\overline{\mathrm{Cov}}_{\tilde{\boldsymbol{0}}}(\hat{\mathcal{N}}_{A_{x_i}},\hat{\mathcal{N}}_{B_{x_j}})=\mathrm{Cov}_{\tilde{\boldsymbol{0}}}(\hat{\mathcal{N}}_{A_{x_i}},\hat{\mathcal{N}}_{B_{x_j}})/\sqrt{\langle \hat{\mathcal{N}}_{A_{x_i}}\rangle_{\tilde{\boldsymbol{0}}} \langle \hat{\mathcal{N}}_{B_{x_j}}\rangle_{\tilde{\boldsymbol{0}}}}$.
Such a normalization gives $\overline{\mathrm{Cov}}_{\tilde{\boldsymbol{0}}}(\hat{\mathcal{N}}_{A_{x_i}},\hat{\mathcal{N}}_{A_{x_i}})=1$ in the case of a Poissonian distribution for excitations in a single QDO. For all the atomic QDOs in the complex, the values of the atomic mean excitation number are below $10^{-2}$, suggesting that the dipolar interactions in the MBD ground state act as a perturbation on
the uncoupled atomic QDO system. Such an observation is confirmed
by the strength of normalized correlations between QDOs not exceeding 0.2.
The QDOs on the fullerene have
higher mean excitation numbers, and their mutual correlations are stronger 
compared with the atomic QDOs associated to the carbon atoms in the CPPA molecule. This suggests that collective effects are stronger in a compact quasi-spherical homonuclear fullerene. The asymmetry of the fullerene position with respect to the CPPA ring manifests itself in
an enhancement of the excitation of the $C_{70}$ QDOs located closer to the phenyl rings. This can be interpreted as a polarization effect on the fullerene due to the CPPA acting as an external environment. 
Similar conclusions are reached when the fullerene is considered as an external
environment acting on the CPPA.
Excitation number analysis in the SQ-MBD framework can be easily extended to more complex systems, 
providing a useful tool to investigate the effect of a general environment on coupled QDOs, thus paving the way for the development of effective models of MBD interactions in open systems. 
\begin{figure*}[tbh!]
    \centering
      \includegraphics[scale=0.0835]{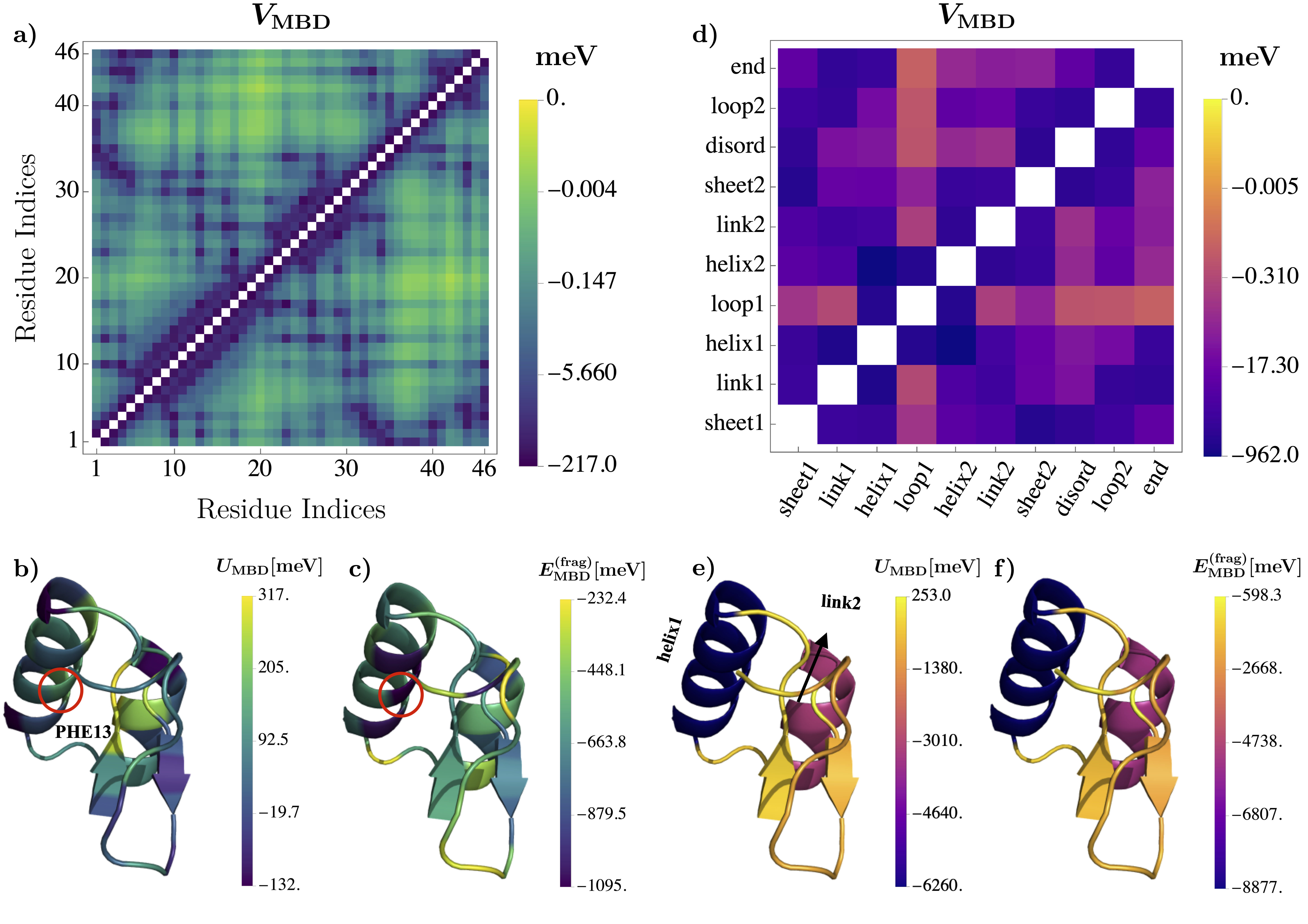}
  \caption{\textbf{Contributions to the MBD energy for the crambin protein [PDB ID: 2FD7] from two different
  coarse-grained partitions (fragments) of atomic QDOs}. $V_{\rm MBD}$ is the MBD mutual interaction energy between fragments, $U_{\rm MBD}$ is the internal MBD energy of each fragment, and $E_{\rm MBD}^{(\rm frag)}$ is the total fragment contribution to the MBD energy. Left panel corresponds to residue fragments, whereas right panel corresponds to the specified secondary-structure fragments.}
      \label{fig:EMBD_reproj}
\end{figure*}

The SQ-MBD framework also
considerably simplifies the calculation
of operator expectation values between fragments in the collective MBD state. 
Let us consider a partition $\{\mathcal{F}_{\alpha}\}_{\alpha=1}^{N_{\rm frag}}$ of the whole system, $\mathcal{S}=\cup_{\alpha}\mathcal{F}_{\alpha}$,
each fragment being specified by a set of atomic QDOs, $\mathcal{F}_{\alpha}=\{A_{1},...,A_{N_{\alpha}}\}$.
For such a partition, it is possible to define the single- and pair-fragment contributions to the total MBD energy of the system $E_{\rm}$,
\begin{equation}
E_{\text{MBD}} = \sum_{\alpha,\beta=1}^{N_{\rm frag}} (E_{\rm MBD})_{\alpha\beta}= \sum_{\alpha =1}^{N_{\rm frag}} (E^{(\rm frag)}_{\rm MBD})_{\alpha}
\end{equation} 
where
$
(E_{\rm MBD})_{\alpha \alpha}=(U_{\rm MBD})_{\alpha}=\langle \tilde{\boldsymbol{0}}|\hat{H}_{\rm MBD}|_{\mathcal{F}_{\alpha}} |\tilde{\boldsymbol{0}}\rangle-(\hbar/2)\sum_{A\in\mathcal{\alpha}} \omega_{A}
$
is the internal MBD energy of the $\alpha$-th fragment,
$
(E_{\rm MBD})_{\alpha\beta}=(V_{\rm MBD})_{\alpha \beta}= 1/2 \times
\langle \tilde{\boldsymbol{0}}|\hat{H}_{\rm MBD}|_{\mathcal{F}_{\alpha}\cup\mathcal{F}_{\beta}} |\tilde{\boldsymbol{0}}\rangle
$
is the mutual MBD interaction energy between the $\alpha$-th and 
$\beta$-th fragments with $\alpha \neq \beta$,  and $(E^{(\rm frag)}_{\rm MBD})_{\alpha}=\sum_{ \beta=1}^{N_{\rm frag}} (E_{\rm MBD})_{\alpha\beta}$ is the total contribution of the $\alpha$-th 
fragment to the total MBD energy. We stress that such quantification of the fragment contribution to the collective MBD energy is not unique since there are multiple ways to partition the pair-fragment contributions. 
However, the proposed fragment-based projection scheme can be straightforwardly applied to large molecular systems with arbitrary levels of coarse-graining \cite{marrink2007martini,ingolfsson2014power,yu2021multiscale}.
As a case study, here we consider crambin (see Figure 
\ref{fig:EMBD_reproj}), a protein with 46 amino acid residues exhibiting essentially all relevant secondary-structure motifs and that has 
been extensively used as a model for crystallography, NMR technique development, and folding studies \cite{bang2004total}. 
The energy scale of single-residue contributions to the MBD energy is $|(E^{(\rm frag)}_{\rm MBD})_{\alpha}|\sim 0.1-1 \,\,\rm{eV}$, while for larger secondary-structure elements $|(E^{(\rm frag)}_{\rm MBD})_{\alpha}|\sim 0.5-9 \,\,\rm{eV}$ -- as strong as covalent bonds. This reinforces the relevance of dispersion interactions and their interplay with covalent bonding in driving the dynamics of biomolecular systems. 
Interestingly, there are fragments exhibiting a \emph{positive} internal MBD energy both in the case of residues (see Figure 
\ref{fig:EMBD_reproj}b) and of secondary structures (see Figure \ref{fig:EMBD_reproj}e). In particular, the residues with a positive internal
MBD energy represent a large majority (33 of 46), while the only secondary structure motif with positive $(U_{\rm MBD})_{\alpha}$ is the link2 structure.
This effect can be interpreted as the screening of the intrafragment MBD interactions due to the presence of the external environment, 
leading to a blueshift of atomic QDO frequencies not compensated by the negative energy contribution due to the mutual dipolar interactions
between QDOs inside the fragment.
However, the total single-fragment contribution to the MBD energy for all the fragments in the biomolecule is negative for the considered coarse-graining schemes
(see Figure \ref{fig:EMBD_reproj}a and Figure 
\ref{fig:EMBD_reproj}d). 
Particularly suggestive is the case of the phenylalanine residue (PHE13) located at the
center of the longest alpha helix: it is a fragment with a high positive internal MBD energy ($(U_{\rm MBD})_{\rm PHE13}\sim 0.13 \,\,\rm{eV}$), and it is also the residue with the largest negative single-fragment energy contribution  ($(E_{\rm MBD}^{\rm (frag)})_{\rm PHE 13}\sim -1.1 \,\,\rm{eV}$). This can be interpreted as a fingerprint of the strong coupling of the fragment with the external environment: the correlations among atomic QDOs inside the fragment are disturbed in favor of establishing stronger correlation with the rest of the protein.
On the other hand, the difference between single-fragment contributions to MBD and the internal MBD energy for the
longest helix in the complex is $\sim -2.2 \,\,\rm{eV}$
compared with $(U_{\rm MBD})_{\rm helix1}\sim -6.3\,\, \rm{eV}$. This means that the QDOs inside the alpha-helix are strongly correlated and interacting, constituting a fragment weakly coupled to the rest of the protein. This analysis suggests a possible strategy to develop a 
coarse-grained model of MBD interactions similar to existing 
quantum embedding methods \cite{knizia2012density,sun2016quantum,lin2021variational,ma2021quantum}: identifying the fragments with stronger internal correlations and interactions among atomic QDOs, 
solving the coupled QDOs inside these fragments and treating the weaker interactions with the rest of the system
in a perturbative way, in analogy with the orbital hybridization description of covalent interactions. 
The  $N_{\rm frag}\times N_{\rm frag}$ matrix $V_{\rm MBD}$ and 
the $N_{\rm frag}$-dimensional vectors $U_{\rm MBD},E^{(\rm frag)}$
provide low-rank representations of MBD interactions and could be used to develop these coarse-grained models, which can serve as inputs to machine-learned force fields~\cite{chmiela2019sgdml,noe2020machine,musil2021physics}.

In addition, the SQ-MBD analysis allows one to decompose the interaction energy contained in the fully coupled MBD state of the whole system without the need for arbitrary projections into states of isolated fragments, as it is usually done in first-quantized MBD calculations~\cite{hermann2017nanoscale,stohr2019quantum}.
As we show in section IX of the Supplemental Material, the MBD interaction energy of a molecular complex can be decomposed into ``bonding,'' ``anti-bonding,'' and a background of ``weakly interacting'' modes (or ``non-covalent orbitals''). Furthermore, the MBD eigenvectors associated
with the ``bonding'' modes appear as a ``solenoidal'' vector field
over the atoms in the complex. In contrast, the MBD eigenvectors associated with ``anti-bonding'' modes exhibit an ``irrotational'' vector field. This insight into normal modes and their contribution to the binding energy could be used to develop advanced coarse-graining procedures. Hence, SQ-MBD offers a clear advantage over its first-quantized counterpart in terms of identifying non-covalent orbitals that determine interactions in large chemical and biological systems. 

Finally, we demonstrate insights into electronic and quantum-information properties enabled by the SQ-MBD method. MBD transition dipoles are relevant quantities required to calculate the static and dynamic polarizabilities of the coupled QDOs.
\begin{figure*}[tbh!]
        \centering
    	\includegraphics[scale=0.0845]{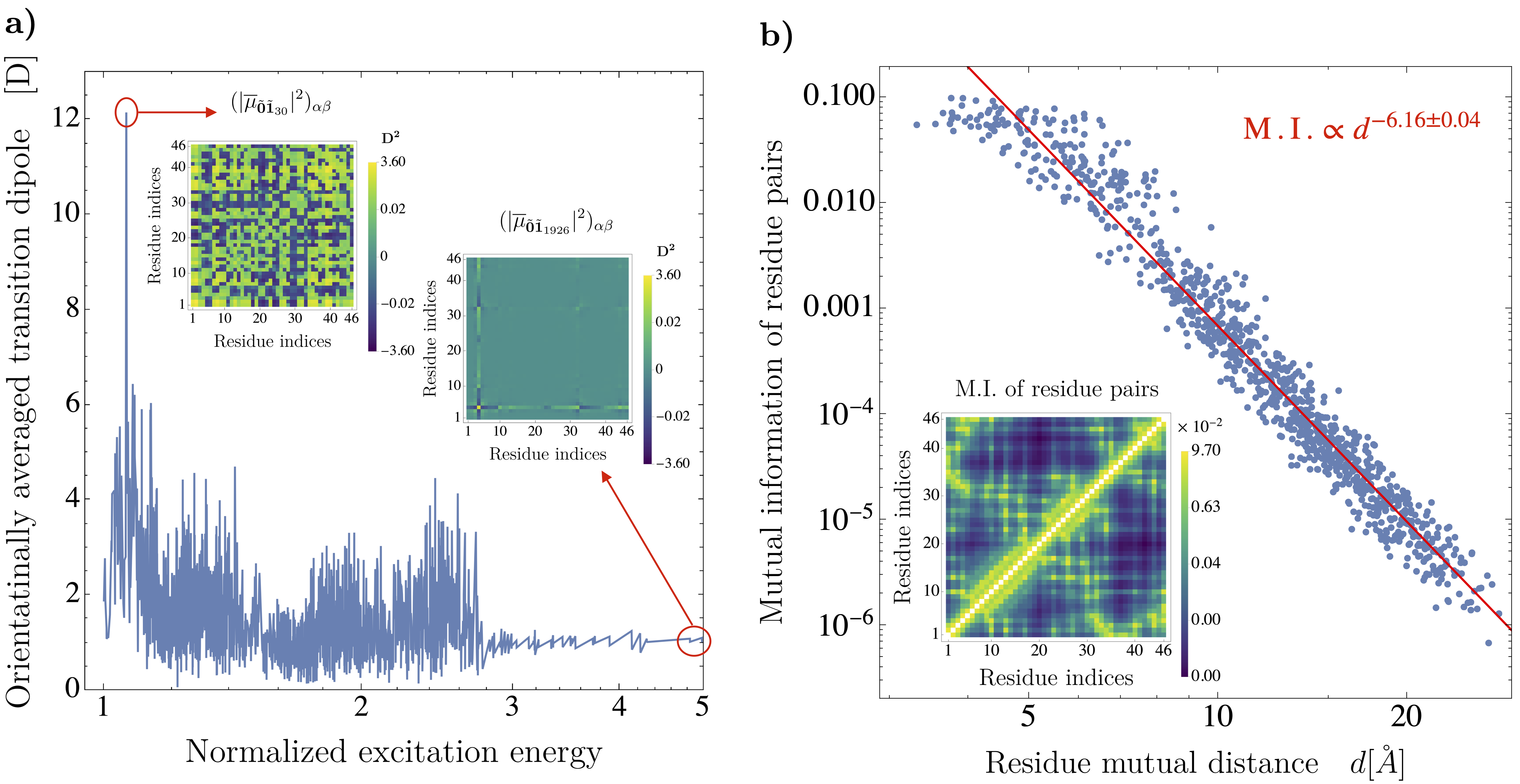} 
  \caption{\textbf{Transition dipoles of MBD modes and quantum-information analysis of the MBD ground state for crambin residues}.
  Left panel (a) shows the MBD transition dipole $|\bar{\mu}_{\tilde{\boldsymbol{0}}\tilde{\boldsymbol{1}}_k}|$ vs. the normalized MBD excitation energy  $\hbar \tilde{\omega}_{k}/(\hbar \tilde{\omega}_1)$. In the insets, the matrix elements of the square of the orientationally averaged transition dipole for interacting residues 
$
(|\bar{\mu}_{\tilde{\boldsymbol{0}}\tilde{\boldsymbol{1}}_k}|^2)_{\alpha \beta}=
1/3 \sum_{A\in\mathcal{F}_{\alpha}, B\in\mathcal{F}_{\beta}}\sum_{i=1}^3\,\mu_{\tilde{\boldsymbol{0}}\tilde{\boldsymbol{1}}_k;A_{x_i}}\mu^{*}_{\tilde{\boldsymbol{0}}\tilde{\boldsymbol{1}}_k;B_{x_i}}$ are shown. Right panel (b) shows the scatter plot of the mutual information (M.I.) for pairs of residues vs. the distance ($d$) between their centers of mass. In the inset, the mutual information of residue pairs is shown.}
\label{fig:AppIIQformalism}
\end{figure*}
Due to Fock-state selection rules, transition dipoles in a system of linearly coupled QDOs are allowed only between the MBD ground state 
$|\tilde{\boldsymbol{0}}\rangle $ and singly-excited states,
$|\tilde{\boldsymbol{1}}_k\rangle=|\tilde{0}_1,...,\tilde{1}_{k},...\tilde{0}_{3N}\rangle$.
The $x_i$-th Cartesian component of the
electric transition dipole between the
ground state and the state with a single excitation in the $k$-th MBD mode is given by
$
    (\mu_{x_i})_{\tilde{\boldsymbol{0}}\tilde{\boldsymbol{1}}_k}=\sum_{A=1}^N \langle \tilde{\boldsymbol{1}}_k | \hat{\mu}_{A_{x_i}} | \tilde{\mathbf{0}} \rangle\,.
$
We introduce the scalar quantity 
\begin{equation}
\label{eqref:isodip_kmode}
|\bar{\mu}_{\tilde{\boldsymbol{0}}\tilde{\boldsymbol{1}}_k}|^2=\dfrac{1}{3}\sum_{A,B=1}^N \sum_{i=1}^3 \mu_{\tilde{\boldsymbol{0}}\tilde{\boldsymbol{1}}_k;A_{x_i}}\mu^{*}_{\tilde{\boldsymbol{0}}\tilde{\boldsymbol{1}}_k§;B_{x_i}}
\end{equation}
that we will refer to as the isotropic (orientationally averaged) square modulus of the transition dipole associated to the $k$-th MBD normal mode and that can be interpreted as the contribution of a specific MBD normal mode to the isotropic polarizability of the system. In Figure \ref{fig:AppIIQformalism}a, the plot of $|\bar{\mu}_{\tilde{\boldsymbol{0}}\tilde{\boldsymbol{1}}_k}|$ 
as a function of the normalized eigenenergy $\hbar \tilde{\omega}_{k}/(\hbar \tilde{\omega}_1)$ (adimensionalized by the lowest MBD eigenenergy units) associated to the $k$-th MBD normal mode for crambin is shown.
The MBD normal modes in the range between $1-2.5\,\,\tilde{\omega}_{1}$ exhibit a rather high isotropic transition dipole moment $\sim 6-8\,\, \mathrm{D}$ on average, of the same order of magnitude as the transition dipoles of organic 
fluorophores \cite{pearlstein1991theoretical,chung2016determining,kurian2017oxidative,celardo2019existence}. A single mode near $ 1.06 \,\,\tilde{\omega}_1$ has a much larger transition dipole $\sim 12\,\,\mathrm{D}$,
arising from a strong collective atomic response. 
In many practical applications, it is relevant to ascertain how strongly
correlated are transition dipole elements between specific
residues. With this aim in mind, we introduce 
the two-fragment contribution to the effective square modulus of the isotropic transition dipole.
In the insets of Figure \ref{fig:AppIIQformalism}a, plots of the matrix defined in Eq. \eqref{eqref:isodip_kmode} are reported for two MBD normal modes: the low-frequency mode with the highest transition 
dipole ($k=13$) and the highest-frequency MBD mode ($k=1926$) having a much smaller total transition dipole.
The results show that the low-frequency mode strongly correlates dipole fluctuations over many residues of the whole molecule, while the high-frequency mode correlates fluctuations essentially of a single residue. The previous considerations and analysis can  be extended to the calculation of higher-order transition multipoles. 
In the context of coarse graining MBD calculations, the SQ-MBD framework allows access to mutual quantum information among atomic QDOs arising from MBD normal modes. 
The application of quantum information methods to study electronic
correlation properties has recently found successful application
in the construction of a new class of correlation energy functionals
in reduced density matrix functional theory \cite{wang2021information,wang2022self}.
In the MBD framework, the mutual information can be used to partition a given system into fragments that minimize inter-fragment MBD interactions.
In particular, the unitary representation of 
Bogoliubov transformations in Eq.\eqref{eqref:Gs_QDOtoMBD}
shows that the MBD ground state is a multimodal Gaussian state of
the same type as the ones used in continuous-variable
quantum information theory
\cite{weedbrook2012gaussian,adesso2014continuous}.
The quantum information-derived observable quantifying correlations among
two parts of a given quantum system is the mutual information defined as
\begin{equation}
    (\mathrm{M.I.})_{\alpha \beta} = S[\hat{\rho}_{\alpha}]+S[\hat{\rho}_{\beta}]-S[\hat{\rho}_{\alpha \beta}],
\end{equation}
where $S[\rho_{\alpha}]=-\mathrm{Tr}[\hat{\rho}_{\alpha}\log \hat{\rho}_{\alpha}]$ is the von Neumann entropy of the reduced density matrix $\hat{\rho}_{\alpha}=\mathrm{Tr}_{\gamma
\neq \alpha} (|\tilde{\boldsymbol{0}}\rangle\langle \tilde{\boldsymbol{0}}|)$
and $\hat{\rho}_{\alpha\beta}=\mathrm{Tr}_{\gamma
\neq \alpha,\beta} (|\tilde{\boldsymbol{0}}\rangle\langle \tilde{\boldsymbol{0}}|)$.
The method used to evaluate the von Neumann entropy for a fragment $\mathcal{F}_{\alpha}$ relies on the symplectic spectrum of the correlation matrix $\Sigma^{(\alpha)}$
\begin{equation}
\Sigma^{(\alpha)}=
\begin{pmatrix}
\langle \hat{\boldsymbol{a}}_{\alpha}\otimes \hat{\boldsymbol{a}}_{\alpha} \rangle_{\tilde{\mathbf{0}}\tilde{\mathbf{0}}} && \langle \hat{\boldsymbol{a}}_{\alpha}\otimes \hat{\boldsymbol{a}}_{\alpha}^{\dagger} \rangle_{\tilde{\mathbf{0}}\tilde{\mathbf{0}}}\\
\langle \hat{\boldsymbol{a}}_{\alpha}^{\dagger}\otimes \hat{\boldsymbol{a}}_{\alpha} \rangle_{\tilde{\mathbf{0}}\tilde{\mathbf{0}}} && \langle \hat{\boldsymbol{a}}_{\alpha}^{\dagger}\otimes \hat{\boldsymbol{a}}_{\alpha}^{\dagger} \rangle_{\tilde{\mathbf{0}}\tilde{\mathbf{0}}}
\end{pmatrix}
\end{equation}
(see \cite{weedbrook2012gaussian,adesso2014continuous} and Supplemental Material for further details). In the inset of Figure \ref{fig:AppIIQformalism}b, the results of the calculation of mutual information between pairs of residues in the MBD ground state are reported. As expected, the mutual information between residues is strongly correlated with the inter-residue distance $d$. Figure~\ref{fig:AppIIQformalism}b reveals that $\text{M.I.}$ scales as $\sim d^{-6.16\pm 0.04}$, albeit with a substantial scatter. This scaling law follows the inter-residue vdW interaction energy; in fact, the mutual information
between pairs of residues and the mutual interaction energies $V_{\rm MBD}$ for the 
same pairs are strongly correlated (see Supplemental Material). This suggests that mutual information between fragments and mutual interaction energy $V_{\rm MBD}$ carry similar information about the correlations among QDOs. A more advanced analysis can be developed in the same SQ-MBD framework, examining multifragment correlations among QDOs and generalizing mutual information concepts to multipartite-entangled systems~\cite{ou2007monogamy, gori2021theory}.

In summary, we have presented a formulation of the MBD model in the second quantization picture (SQ-MBD), leading to novel computational and conceptual insights into coupled QDOs in intricate molecular systems. The presented method allowed us to investigate the ground state of the MBD Hamiltonian in terms of the superposition of QDO excited states. Owing to the Fock space representation in the SQ-MBD framework, it becomes possible to simplify the calculation of expectation values of observables in and between MBD ground and excited states. SQ-MBD thus provides a suitable framework to compute and analyze the contribution of arbitrary sub-fragments to many important properties
of the whole system, including the total MBD energy or the transition dipoles of MBD modes.
The SQ-MBD approach could be 
extended to periodic systems akin to its parent MBD Hamiltonian~\cite{Hermann-CR}, where
cooperative effects among atomic QDOs (mediated by 
plasmon-like MBD modes) may provide enhanced insights owing to the high symmetry of Bloch states in such systems.
Moreover, SQ-MBD provides a natural approach to connect the MBD model of coupled QDOs with quantum information theory, paving the way to a straightforward application of methods to analyze correlations and (non)separability among different fragments in complex molecular systems. These results thus represent the starting point for the development of computationally efficient strategies to enable the application of MBD interactions to molecular complexes with millions of atoms.\\

MG and AT acknowledge support from the European Research Council (ERC Consolidator Grant ``BeStMo'') and Fonds National de la Recherche Luxembourg (FNR CORE grant BroadApp C20/MS/14769845). MG would like to thank Mario Galante, Marco Pezzutto, and Martin St\"ohr for their helpful suggestions. PK acknowledges support from the National Science Foundation, Whole Genome Science Foundation, and the U.S.-Italy Fulbright Commission. The authors would also like to acknowledge discussions at the Institute for Pure and Applied Mathematics, and insights from Georgia Dunston, Marco Pettini, and Giuseppe Vitiello. 

\bibliographystyle{apsrev4-2}
\bibliography{apssamp.bib}
\end{document}


\preprint{APS/123-QED}

\title{Supplemental Material for ``Second Quantization Approach to Many-Body Dispersion Interactions: Implications for Chemical and Biological Systems'' }

\author{Matteo Gori}
\affiliation{Department of Physics and Materials Science, University of Luxembourg, L-1511 Luxembourg City, Luxembourg}
 \affiliation{Quantum Biology Laboratory, Howard  University, Washington DC 20060, USA}
 \author{Philip Kurian}
 \affiliation{Quantum Biology Laboratory, Howard  University, Washington DC 20060, USA}
\author{Alexandre Tkatchenko}
\affiliation{%
 Department of Physics and Materials Science, University of Luxembourg, L-1511 Luxembourg City, Luxembourg}

\maketitle

\section{Details of the many-body dispersion (MBD) model}

The parametrization of quantum Drude oscillators (QDOs) adopted throughout this paper is
based on a two-step procedure.
First, the polarizability $\mathcal{A}_A$ (expressed
in volume units), the $C_{6,AA}$ dispersion coefficient, and the van der Waals radius $R_{vdW,A}$ of each atom $A$ in the system were calibrated via the
Tkatchenko-Scheffler method \cite{tkatchenko2009accurate},
\begin{equation}
\begin{cases}
    &C^{(aim)}_{6,AA}= \eta_A^{2} C^{(0)}_{6,AA}\\
    & \mathcal{A}^{(aim)}_A = \eta_A \mathcal{A}^{(0)}_{A}\\
    & \mathcal{R}^{(aim)}_{vdW,A} = \eta_A^{1/3} R^{(0)}_{vdW,A}
\end{cases}
\end{equation}
where $\eta_A=h_{A}/Z_A$ is the ratio \footnote{
The estimation of the $\eta_A$ ratios is a key step in setting the  parameters of the atomic QDOs. In order to refine the estimation of
non-local effects for atom-in-molecule (AIM) response properties, the natural orbital functional theory framework could provide an alternative choice, taking into account the contribution to AIM electric response properties of the off-diagonal density matrix elements} between the
on-site contribution $h_{A}$ to the Mulliken population (corresponding to the atom-projected trace of the density matrix) and the atomic charge $Z_A$ in the case of a free atom \cite{stohr2016communication} applied to electronic structure calculations from density functional-based tight binding (DFTB) using a modified version of the software package DFTB+ \cite{hourahine2020dftb+}.
For the second step, the diagonalization of the MBD Hamiltonian was executed using the software library LibMBD \cite{libmbd}.
In this step of the parametrization,
the $C_{6,AA}^{(aim)}$ coefficients and the 
polarizabilities $\mathcal{A}^{(aim)}_A$ are rescaled according to the
``range-separated self-consistent screening"  procedure (the `rsscs' option
in LibMBD) to account for electrodynamic screening
using the short-range part of the range-separated dipole tensor for 
quantum harmonic oscillators (see \cite{stohr2019theory}). In this way, the parameters $\mathcal{A}_A,C_{6,AA},R_{vdW,A}$ of each 
 atom $A$ have been fixed, taking into account both the 
atom-in-molecule features and the self-consistent screening.

The dipole-dipole potential between a pair of three-dimensional QDOs is given by the rank-2 ($3\times 3$) tensor 
\begin{equation}
   \mathbb{T}_{AB}(\mathbf{R}_{AB})=\nabla_{\mathbf{R}_A}\otimes \nabla_{\mathbf{R}_B} \left(\dfrac{1}{|\mathbf{R}_{AB}|}\right)=\dfrac{(\boldsymbol{I}-3\hat{\boldsymbol{R}}_{AB}\otimes\hat{\boldsymbol{R}}_{AB})}{|\mathbf{R}_{AB}|^3}
\end{equation}
where $\boldsymbol{R}_{AB}=\boldsymbol{R}_{B}-\boldsymbol{R}_A$ is the
distance vector between atoms $A$ and $B$. Adopting the notation presented in \cite{stohr2019theory}, the damping function for each atomic oscillator pair's range-separated potential was chosen
to be
\begin{equation}
    g_{rs,AB}(\|\mathbf{R}_{AB}\|)= 1-\dfrac{1}{1+\exp\{-a[\|\mathbf{R}_{AB}\|/(\beta R^{(aim)}_{vdW,AB})-1]\}},
\end{equation}
where $R^{(aim)}_{vdW,AB}=R^{(aim)}_{vdW,A}+R^{(aim)}_{vdW,B}$ and the parameters $a=6.0$ and $\beta = 0.83$ have been fixed.
The characteristic angular frequency of the atom $A$ is given by 
\begin{equation}
    \omega_A = \dfrac{4}{3}\dfrac{C_{6,AA}}{\hbar\mathcal{A}^2_{A}}\,.
\end{equation}
The diagonalization of the $3N \times 3N$ MBD Hamiltonian matrix
was performed using LibMBD software, thus obtaining the orthogonal transformation $\mathcal{O}$ between atomic QDOs and MBD normal modes, as well as the $3N$ eigenvalues $\tilde{\omega}_k$
corresponding to these MBD eigenmode frequencies.

\section{Bogoliubov transformation for quantum Drude oscillators (QDOs) in the SQ-MBD framework}

The matrix $M(\boldsymbol{\omega})$ maps from the first- to second-quantization representation for the atomic QDOs and takes the form
\begin{equation}
    M(\boldsymbol{\omega})=
    \begin{pmatrix}
    M_{\boldsymbol{a}\boldsymbol{q}} & M_{\boldsymbol{a}\boldsymbol{p}} \\
   M_{\boldsymbol{a^{\dagger}}\boldsymbol{q}} &
   M_{\boldsymbol{a^{\dagger}}\boldsymbol{p}} \\ 
    \end{pmatrix}=
    \dfrac{1}{\sqrt{2\hbar}}
    \begin{pmatrix}
    \mathcal{D}^{1/2}(\boldsymbol{\omega}) & \mathrm{i}\mathcal{D}^{-1/2}(\boldsymbol{\omega})\\
    \mathcal{D}^{1/2}(\boldsymbol{\omega}) & -\mathrm{i}\mathcal{D}^{-1/2}(\boldsymbol{\omega})\\
    \end{pmatrix},
\end{equation}
where $\mathcal{D}(\boldsymbol{\omega})=\mathrm{diag}(\omega_1,\omega_1,\omega_1,\omega_2,\omega_2, \omega_2,...,\omega_{N},\omega_{N},\omega_{N})$ 
is a $3N \times 3N$ diagonal matrix of QDO eigenfrequencies. $\tilde{M}(\tilde{\boldsymbol{\omega}}$) maps from first- to second-quantization representation for the MBD normal modes.
The X,Y matrices in Eq. (2) of the main text define the linear transformation between 
the creation/annihilation operator algebra for the atomic QDOs and the one for the
MBD normal modes. These matrices can be expressed in terms of the orthogonal matrices $\mathcal{O}$ and the transformation matrices $M(\boldsymbol{\omega}),\tilde{M}(\tilde{\boldsymbol{\omega}})$:
 \begin{equation}
\label{eq:bogo_TotMatrix_XY}
\begin{split}
  & X =\left[\tilde{M}_{\boldsymbol{b}\tilde{\boldsymbol{q}}}(\tilde{\boldsymbol{\omega}})\mathcal{O}_{\tilde{\boldsymbol{q}}\boldsymbol{q}}M_{\boldsymbol{q}\boldsymbol{a}}(\boldsymbol{\omega})+\tilde{M}_{\boldsymbol{b}\tilde{\boldsymbol{p}}}(\tilde{\boldsymbol{\omega}})\mathcal{O}_{\tilde{\boldsymbol{p}}\boldsymbol{p}}M_{\boldsymbol{p}\boldsymbol{a}}(\boldsymbol{\omega})\right]=\dfrac{1}{2}\left[\tilde{\mathcal{D}}^{1/2}\mathcal{O}\mathcal{D}^{-1/2}+\tilde{\mathcal{D}}^{-1/2}\mathcal{O}\mathcal{D}^{1/2}\right]\\
  \\
  & Y=\left[\tilde{M}_{\boldsymbol{b}\tilde{\boldsymbol{q}}}(\tilde{\boldsymbol{\omega}})\mathcal{O}_{\tilde{\boldsymbol{q}}\boldsymbol{x}}M_{\boldsymbol{q}\boldsymbol{a}}(\boldsymbol{\omega})-\tilde{M}_{\boldsymbol{b}\tilde{\boldsymbol{p}}}(\tilde{\boldsymbol{\omega}})\mathcal{O}_{\tilde{\boldsymbol{p}}\boldsymbol{p}}M_{\boldsymbol{p}\boldsymbol{a}}(\boldsymbol{\omega})\right]=\dfrac{1}{2}\left[\tilde{\mathcal{D}}^{1/2}\mathcal{O}\mathcal{D}^{-1/2}-\tilde{\mathcal{D}}^{-1/2}\mathcal{O}\mathcal{D}^{1/2}\right]\,\,.
\end{split}
\end{equation}
It can be easily checked that such $3N \times 3N$ real matrices $X,Y$ satisfy the Bogoliubov conditions
assuring preservation of the canonical commutation relations for the sets of
creation/annihilation operators \cite{ripka1986quantum},
\begin{equation}
\label{eq:BogCond}
\begin{split}
&X X^{\dagger}-YY^{\dagger}=\mathbb{I}_{3N\times 3N}, \qquad YX^{\text{T}}-XY^{\text{T}}=0\\
&X^{*} X^{\text{T}}-Y^{*}Y^{\text{T}}=\mathbb{I}_{3N\times 3N}, \qquad  Y^{*}X^{\dagger}-X^{*}Y^{\dagger}=0\,\,.
\end{split}
\end{equation}
The conditions in the first and second lines of Eq. (\ref{eq:BogCond}) are equivalent in the case of real matrices. In fact,
\begin{equation}
\begin{split}
   & XX^{\rm T}-YY^{\rm T}=\dfrac{1}{4}\biggr[(\tilde{\mathcal{D}}^{1/2}\mathcal{O}\mathcal{D}^{-1/2}+\tilde{\mathcal{D}}^{-1/2}\mathcal{O}\mathcal{D}^{1/2})(\mathcal{D}^{-1/2}\mathcal{O}^{\rm T} \tilde{\mathcal{D}}^{1/2}+\mathcal{D}^{1/2} \mathcal{O}^{\rm T} \tilde{\mathcal{D}}^{-1/2})+\\
   & (\tilde{\mathcal{D}}^{1/2}\mathcal{O}\mathcal{D}^{-1/2}-\tilde{\mathcal{D}}^{-1/2}\mathcal{O}\mathcal{D}^{1/2})(\mathcal{D}^{-1/2}\mathcal{O}^{\rm T} \tilde{\mathcal{D}}^{1/2}-\mathcal{D}^{1/2} \mathcal{O}^{\rm T} \tilde{\mathcal{D}}^{-1/2}) \biggr]=\dfrac{1}{4}\biggr[
   \tilde{\mathcal{D}}^{1/2}\mathcal{O}\mathcal{D}^{-1}\mathcal{O}^{\rm T}\tilde{\mathcal{D}}^{1/2}+\\
   &+\tilde{\mathcal{D}}^{1/2}\mathcal{O}\mathcal{O}^{\rm T}\tilde{\mathcal{D}}^{-1/2}+\tilde{\mathcal{D}}^{-1/2}\mathcal{O}\mathcal{O}^{\rm T}\tilde{\mathcal{D}}^{1/2}+\tilde{\mathcal{D}}^{-1/2}\mathcal{O}\mathcal{D}\mathcal{O}^{\rm T}\tilde{\mathcal{D}}^{-1/2}-\tilde{\mathcal{D}}^{1/2}\mathcal{O}\mathcal{D}^{-1}\mathcal{O}^{\rm T}\tilde{\mathcal{D}}^{1/2}+\\
   &+\tilde{\mathcal{D}}^{1/2}\mathcal{O}\mathcal{O}^{\rm T}\tilde{\mathcal{D}}^{-1/2}+\tilde{\mathcal{D}}\mathcal{O}\mathcal{O}^{\rm T}\tilde{\mathcal{D}}^{1/2}-\tilde{\mathcal{D}}^{-1/2}\mathcal{O}\mathcal{D}\mathcal{O}^{\rm T}\tilde{\mathcal{D}}^{-1/2}\biggr]=\dfrac{1}{4} (4\mathbb{I})=\mathbb{I}\\
\end{split}
\end{equation}
and
\begin{equation}
\begin{split}
   &YX^{T}-XY^{T}=\dfrac{1}{4}\biggr[\left(\tilde{\mathcal{D}}^{1/2}\mathcal{O}\mathcal{D}^{-1/2}-\tilde{\mathcal{D}}^{-1/2}\mathcal{O}\mathcal{D}^{1/2}\right)\left(\mathcal{D}^{-1/2} \mathcal{O}^{\rm T} \tilde{\mathcal{D}}^{1/2} -\tilde{\mathcal{D}}^{-1/2}\mathcal{O}^{\rm T}\mathcal{D}^{1/2}\right)+\\
   &-(\tilde{\mathcal{D}}^{1/2}\mathcal{O}\mathcal{D}^{-1/2}+\tilde{\mathcal{D}}^{-1/2}\mathcal{O}\mathcal{D}^{1/2})(\mathcal{D}^{-1/2}\mathcal{O}^{\rm T} \tilde{\mathcal{D}}^{1/2} - \mathcal{D}^{1/2} \mathcal{O}^{\rm T} \tilde{\mathcal{D}}^{-1/2})\biggr]=\dfrac{1}{4}\biggr[\tilde{\mathcal{D}}^{1/2} \mathcal{O} \mathcal{D}^{-1} \mathcal{O}^{\rm T} \tilde{\mathcal{D}}^{1/2}+\\
   &+\tilde{\mathcal{D}}^{-1/2}\mathcal{O}\mathcal{O}^{\rm T} \tilde{\mathcal{D}}^{1/2}
   -\tilde{\mathcal{D}}^{-1/2}\mathcal{O}\mathcal{O}^{\rm T}\tilde{\mathcal{D}}^{1/2}-
   \tilde{\mathcal{D}}^{-1/2}\mathcal{O}\mathcal{D}\mathcal{O}^{\rm T}\tilde{\mathcal{D}}^{-1/2}-\tilde{\mathcal{D}}^{1/2}\mathcal{O}\mathcal{D}^{-1}\mathcal{O}^{\rm T} \tilde{\mathcal{D}}^{1/2}+\\
   &+\mathcal{D}^{1/2}\mathcal{O}\mathcal{O}^{\rm T} \tilde{\mathcal{D}}^{-1/2}-\mathcal{D}^{-1/2}\mathcal{O}\mathcal{O}^{\rm T} \tilde{\mathcal{D}}^{1/2}
  +\tilde{\mathcal{D}}^{-1/2}\mathcal{O}\mathcal{D}\mathcal{O}^{\rm T} \tilde{\mathcal{D}}^{-1/2}\biggr]=0.
\end{split}
\end{equation}
It can also be shown that the inverse Bogoliubov transformation for real X,Y matrices is given by \cite{ripka1986quantum}
\begin{equation}
\label{eq:inversT_bogtr}
\begin{pmatrix}
\hat{\boldsymbol{a}}\\
\hat{\boldsymbol{a}}^{\dagger}\\
\end{pmatrix}
=
\begin{pmatrix}
X^{\rm T} & -Y^{\rm T}\\
-Y^{\rm T} & X^{\rm T}\\ 
\end{pmatrix}
\begin{pmatrix}
\hat{\boldsymbol{b}}\\
\hat{\boldsymbol{b}}^{\dagger}\\
\end{pmatrix}.
\end{equation}

\section{$\mathrm{C}_{70}$ fullerene in [8]-CPPA  cycloparaphenyl ring model}
The supramolecular system of $\mathrm{C}_{70}$ fullerene surrounded by a 
cycloparaphenyl ring composed of 8 units consists of 166 atoms. DFTB+ simulations to calculate the ratios $\eta_A$ were
performed in cluster geometry (no periodic boundary conditions), with values for the 
net charge $Z_{tot}=0$ and Fermi temperature $T=0.001\,\,\mathrm{Ha}$.\\

\section{SQ-MBD reprojection scheme}
The expectation value of the harmonic oscillator energy of atom $A$ in the MBD ground state (for which $\hat{b}_{k}| \tilde{\boldsymbol{0}}\rangle = 0$) is given by 
\begin{equation}
\begin{split}
    &(U_{\rm MBD})_{A}=\sum_{i=1}^{3} \hbar \omega_A \langle  \tilde{\boldsymbol{0}}|\hat{a}_{A_{x_i}}^{\dagger} \hat{a}_{A_{x_i}}| \tilde{\boldsymbol{0}}\rangle= \sum_{i=1}^3 \sum_{k,k'=1}^{3N}  \hbar \omega_A(Y^{T})_{A_{x_i} k}
    (Y^{T})_{A_{x_i} k'} \langle  \tilde{\boldsymbol{0}}|\hat{b}_{k} \hat{b}_{k'}^{\dagger}| \tilde{\boldsymbol{0}}\rangle =\\
    &=\hbar \omega_A \sum_{i=1}^3 (Y^{T}Y)_{A_{x_i} A_{x_i}},
\end{split}
\end{equation}
while the expectation value of the interaction potential operator between two atomic
QDOs is given by
\begin{equation}
\begin{split}
\label{eq:VMBD_elem_SQMBD}
   & (V_{\rm MBD})_{AB}= \dfrac{1}{2}\sum_{i,j=1}^{3} \hbar \omega_A \omega_B (\mathbb{T}_{AB})_{ij}
    \langle  \tilde{\boldsymbol{0}}| \hat{\boldsymbol{q}}_A \hat{\boldsymbol{q}}_B | \tilde{\boldsymbol{0}}\rangle =\dfrac{\hbar \omega_A^{1/2} \omega_B^{1/2}}{4} \sum_{i,j=1}^{3} \hbar \omega_A \omega_B (\mathbb{T}_{AB})_{ij} (\hat{a}_{A_{x_i}}+\hat{a}_{A_{x_i}}^{\dagger})\\
    &\times(\hat{a}_{B_{x_j}}+\hat{a}_{B_{x_j}}^{\dagger})=\dfrac{\hbar \omega_A^{1/2} \omega_B^{1/2}}{4} \sum_{i,j=1}^{3}\sum_{k,k'=1}^{3N} (\mathbb{T}_{AB})_{ij} (X_{A_{x_{i}} k}^{T}-Y_{A_{x_{i}}k}^{T})(X_{B_{x_{j}} k}^{T}-Y_{B_{x_{j}}k}^{T})
    \langle\tilde{\boldsymbol{0}}|\hat{b}_{k} \hat{b}_{k'}^{\dagger}| \tilde{\boldsymbol{0}}\rangle=\\
    &=\dfrac{\hbar \omega_A^{1/2} \omega_B^{1/2}}{4} \sum_{i,j=1}^{3} (Q^{T}Q)_{A_{x_i}B_{x_j}} (\mathbb{T}_{AB})_{ij},
\end{split}
\end{equation}
where we have introduced the matrix $Q(X,Y)=X-Y$.

\section{Crambin model}
\begin{figure}
    \centering
    \includegraphics[scale=0.0825]{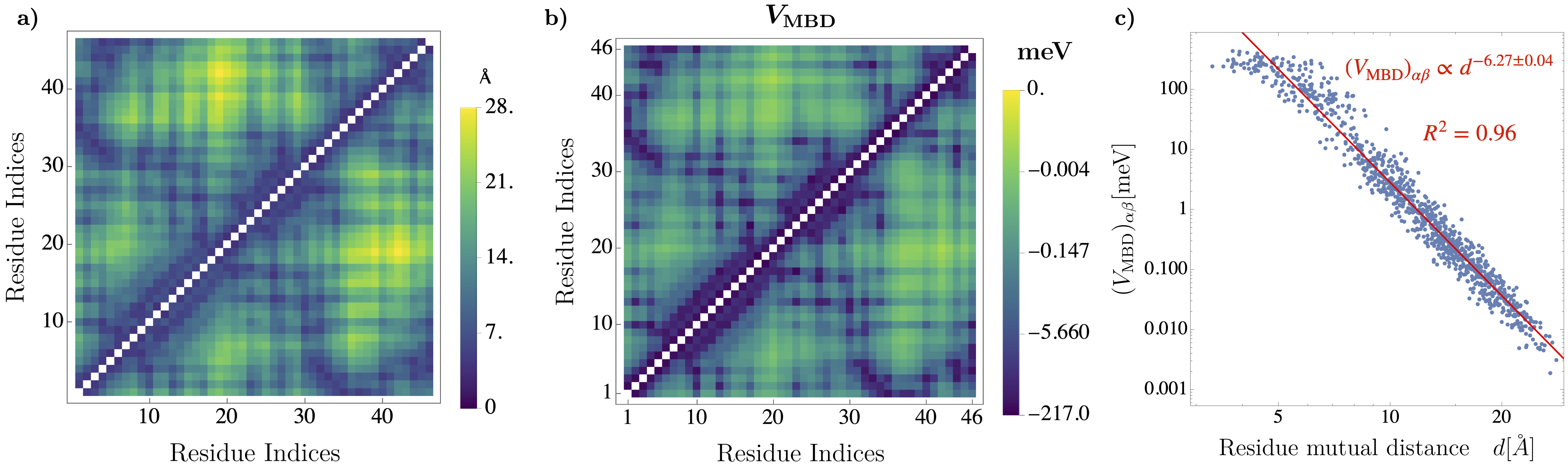}
    \caption{\textbf{Relationship between geometry and mutual MBD energy for pairs of crambin residues}. In panel (a), the distances between geometrical centers of mass for all pairs of residues in crambin are shown. In panel (b), the mutual interaction energies $V_{\rm MBD}$ for all pairs of residues are shown. In panel (c), the correlation plot for the distances between geometrical centers of mass for all pairs of residues and their mutual interaction energies is shown.}
    \label{fig:CrambinReisdues}
\end{figure}

\begin{figure}
    \centering
    \includegraphics[scale=0.0835]{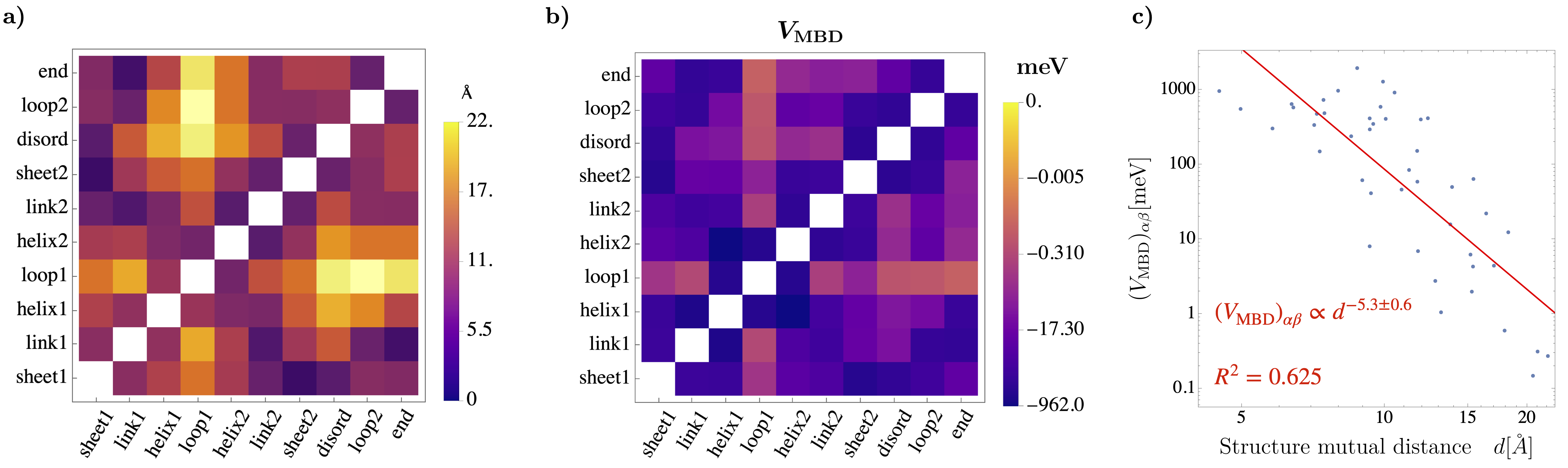}
    \caption{\textbf{Relationship between geometry and mutual MBD energy for pairs of crambin secondary-structure motifs}. In panel (a), the distances between geometrical centers of mass for all pairs of secondary structures in crambin are shown. In panel (b), the mutual interaction energies $V_{\rm MBD}$ for all pairs of secondary structures are shown. In panel (c), the correlation plot for the distances between geometrical centers of mass for all pairs of secondary structures and their mutual interaction energies is shown.}
    \label{fig:CrambinSecondaryStructure}
\end{figure}

The structure of crambin was obtained from Protein Data Bank (PDB) file PDB-ID: 2FD7 and consists of 642 atoms. DFTB+ simulations to calculate the ratios $\eta_A$ were
performed in cluster geometry (no periodic boundary conditions), with values for the 
net charge $Z_{tot}=0$ and Fermi temperature $T=0.001\,\,\mathrm{Ha}$.
To aid interpretation of the $V_{\rm MBD}$ matrices, we report here the distance matrices 
for geometrical barycenters of atomic positions for partitioning of crambin into residues (Figure S\ref{fig:CrambinReisdues})
and into larger secondary structures (Figure S\ref{fig:CrambinSecondaryStructure}).
In the case of residue fragmentation, we can observe
the correlation for the mutual distances between residue geometric centers of mass and
their mutual MBD potential, with the
interactions decaying as $\sim d^{-6.27}$. However, it is interesting to notice a deviation from such
a power law for $d \lesssim \SI{5}{\angstrom}$. This
entails a two-fold explanation. For short length scales, the
metric introduced to measure the distances between residues
becomes less representative if the geometry of the partition, i.e., the reciprocal geometrical configuration of  a pair of fragments, is poorly approximated by the geometrical 
centers of mass. On the other hand, such a deviation could
be attributed to many-body effects among atomic QDOs belonging to the two fragments and with strong mutual correlations.
For the secondary-structure partitioning, it is possible to recognize some
common patterns between the matrices in Figures S\ref{fig:CrambinSecondaryStructure} (a) and (b). However, in this case, the correlation plot reveals a much weaker correlation for the mutual MBD potential and the 
mutual distance between secondary structure motifs.

\section{Transition dipoles in the SQ-MBD framework}
The contribution of atom $A$ along the $x_i$ Cartesian direction to the transition dipole from the MBD ground state $|\tilde{\boldsymbol{0}}\rangle \longrightarrow |\tilde{\boldsymbol{1}}_{k}\rangle$ associated with the
single excitation of the $k$-th MBD mode is given in terms of X,Y matrices by
\begin{equation}
   \mu_{\tilde{\boldsymbol{0}}\tilde{\boldsymbol{1}}_k; A_{x_i}}=\langle\tilde{\boldsymbol{1}}_k | \hat{\mu}_{A_{x_i}} | \tilde{\mathbf{0}} \rangle = 
   \sqrt{\dfrac{\hbar\omega_A\mathcal{A}_{A}}{2}}
   (X-Y)^{T}_{A_{x_i} k}
   \,.
\end{equation}

\section{Quantum information tools for the SQ-MBD model of QDOs}
The MBD ground state as described in the SQ-MBD framework is a multimodal Gaussian bosonic
state. One powerful method to investigate quantum information properties
is based on the Williamson theorem \cite{weedbrook2012gaussian,adesso2014continuous}. Given an algebra of creation/annihilation operators, it is possible to
define the symplectic form
\begin{equation}
\Omega \hat{I}=
\begin{pmatrix}
[\hat{\boldsymbol{a}},\hat{\boldsymbol{a}}] & [\hat{\boldsymbol{a}},\hat{\boldsymbol{a}}^{\dagger}]\\
[\hat{\boldsymbol{a}}^{\dagger},\hat{\boldsymbol{a}}] & [\hat{\boldsymbol{a}}^{\dagger},\hat{\boldsymbol{a}}^{\dagger}]\\
\end{pmatrix}
=
\begin{pmatrix}
0 & \boldsymbol{I}\\
-\boldsymbol{I} & 0 \\
\end{pmatrix}
\hat{I}
\end{equation}
and a group of symplectic transformations preserving the symplectic form such that
\begin{equation}
    \Omega=\mathcal{S}^{T}\Omega \mathcal{S}\,.
\end{equation}
Within this framework, the Williamson theorem states that for any $2N_{\rm frag} \times 2N_{\rm frag}$ covariant matrix $\Sigma^{(\alpha)}$, there exists a symplectic matrix $\mathbf{S}$ such that
\begin{equation}
    \Sigma^{(\alpha)}= \mathbf{S} \Sigma^{(\alpha) \oplus } \mathbf{S}^{\rm T} \qquad  \Sigma^{(\alpha) \oplus }:= \mathrm{diag}(\nu^{\alpha}_1,...,\nu^{\alpha}_{3N})\oplus \mathrm{diag}(\nu^{\alpha}_1,...,\nu^{\alpha}_{3N}),
\end{equation}
where the $N$ positive $\{\nu_k \}_{k=1,..3N}$ quantities are called symplectic 
eigenvalues. They can be easily calculated as the positive eigenspectrum  of the matrix $i\Omega\Sigma^{(\alpha)}$.
The von Neumann entropy associated to the reduced density matrix $\hat{\rho}_{\alpha}=\mathrm{Tr}_{\beta\neq \alpha} |\tilde{\boldsymbol{0}}\rangle\langle \tilde{\boldsymbol{0}}|$
can be calculated by 
\begin{equation}
    S(\hat{\rho}_{\alpha})= \sum_{k=1}^{N_{\rm frag}} f(\nu^{(\alpha)}_{k}),
\end{equation}
where 
\begin{equation}
    f(x)=\left(\dfrac{x+1}{2}\right)\log\left(\dfrac{x+1}{2}\right)-\left(\dfrac{x-1}{2}\right)\log\left(\dfrac{x-1}{2}\right)\,.
\end{equation}

\section{Correlation between potential energy and mutual information for residue pairs}

\begin{figure}[tbh!]
    \centering
    \includegraphics[scale=0.4]{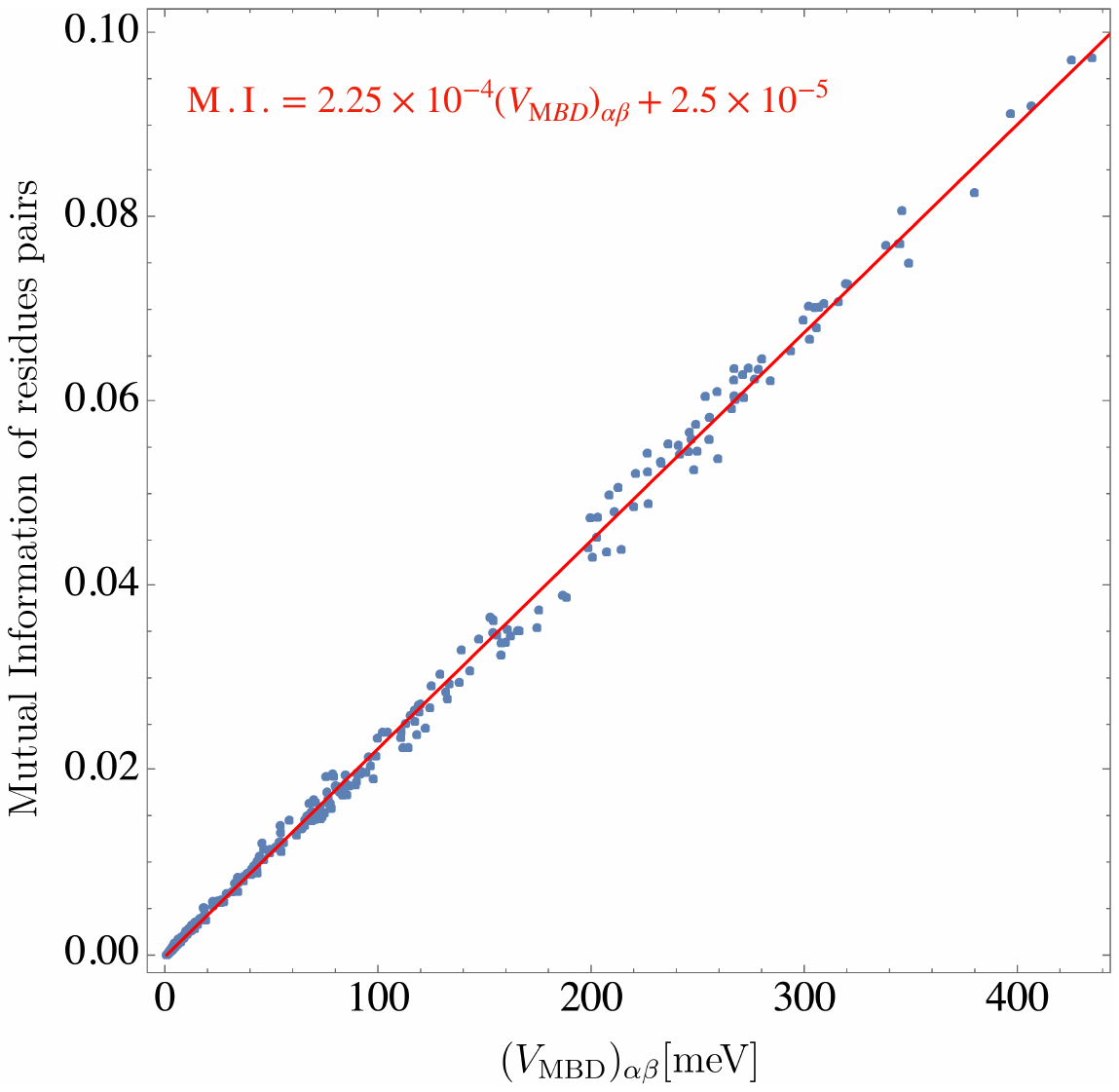}
    \caption{Correlation plot for the mutual information vs. mutual MBD interactions between residue pairs for the crambin protein.}
    \label{fig:CorrVMBDvsMI}
\end{figure}

In order to establish how the mutual MBD interactions between fragment pairs influence their
correlations, we report the correlation plot for the mutual
information $(M.I.)_{\alpha \beta}$ of fragment pairs vs. $(V_{\rm MBD})_{\alpha \beta}$ in Figure S\ref{fig:CorrVMBDvsMI}. It can be noticed how such quantities are strongly correlated and carry the same information on the correlations among QDOs belonging to different fragments.

\pagebreak

\section{ SQ-MBD analysis of the S12L dataset}

The S12L is a dataset of systems stabilized by different kinds of 
supramolecular
interactions and includes dispersion-dominated, $\pi$-stacked,
hydrogen-bonded, and cation–dipolar-bound structures 
\cite{grimme2012supramolecular}.
Such a dataset has been used to benchmark and validate a different
method for estimating van der Waals dispersion energies \cite{grimme2016dispersion}
and for showing how the analysis of MBD plasmon-like modes in 
supramolecular complexes contribute to binding energies \cite{hermann2017nanoscale}.
This section shows how 
the novel SQ-MBD method allows one to decompose information contained in the fully coupled MBD state of the whole system without the need for arbitrary projections into states of isolated fragments, as it is usually done in the first-quantized version of MBD~\cite{hermann2017nanoscale,stohr2019theory}.

\subsection{Analysis of the MBD modes contribution to MBD interaction  between fragments in dispersion-dominated and electrostatically-bound S12L structures}

The SQ-MBD method allows one to easily evaluate the MBD mode
contributions to the expectation value of the quadratic forms of the creation/annihilation operators for the atomic QDOs in the MBD ground state. This analysis reveals the contributions of single QDO excited
states to the MBD ground state and the contribution of the MBD modes
to collective variables.

Let us consider, for instance, the $(\hat{V}_{\rm MBD})_{\mathcal{A}\mathcal{B}}$ operator describing the
interaction between complex $\mathcal{A}$ and complex $\mathcal{B}$ for a 
given system in the S12L dataset in the equilibrium geometry of the complex.
Thanks to the SQ-MBD expression of $(V_{\rm MBD})_{\mathcal{A}\mathcal{B}}=\langle \tilde{\boldsymbol{0}} | (\hat{V}_{\rm MBD})_{\mathcal{A}\mathcal{B}} | \tilde{\boldsymbol{0}} \rangle $ in Eq.~\eqref{eq:VMBD_elem_SQMBD},
it is possible to evaluate the contribution of each MBD mode to the inter-fragment MBD energy. 
We consider the same systems from the S12L database studied
in \cite{hermann2017nanoscale}: a ``tweezer” complex dominated by 
non-polar dispersion interactions in Figure \ref{fig:S12L_2b_geoETmbd}a), and an electrostatically bound
cucurbituril complex in Figure \ref{fig:S12L_7a_geoETmbd}a). In 
that case, the first quantization-based analysis was intended to
identify the displacement component of each atomic QDO particle, giving the
most important contributions to the MBD binding energy $E_{\rm bind,MBD}=(E_{\rm MBD})_{\mathcal{A}\cup\mathcal{B}}-\left[(E_{\rm MBD})_{\mathcal{A}}+(E_{\rm MBD})_{\mathcal{B}}\right]$, i.e., the 
difference between the total MBD energy of the complex and the MBD energy
of the two fragments at an infinite distance. Such an analysis required
the comparison of the given equilibrium configuration with another configuration (fragments at an infinite distance).

Here we propose an alternative observable, the two-fragment contribution
to the MBD potential energy $(V_{\rm MBD})_{\mathcal{A}\mathcal{B}}$, which
is an inherent property of the MBD ground state in a given configuration. Thus, it does not require a comparison with other configurations.
In particular, the SQ-MBD-based analysis we present investigates
the contribution of single MBD modes to the total inter-fragment MBD
energy and the contribution of each pair of atomic QDOs to $(V_{\rm MBD})_{\mathcal{A}\mathcal{B}}$ for a given 
MBD mode.

\subsubsection{The ``tweezer” complex}
\begin{figure}[tbh!]
    \centering
    \includegraphics[scale=0.55]{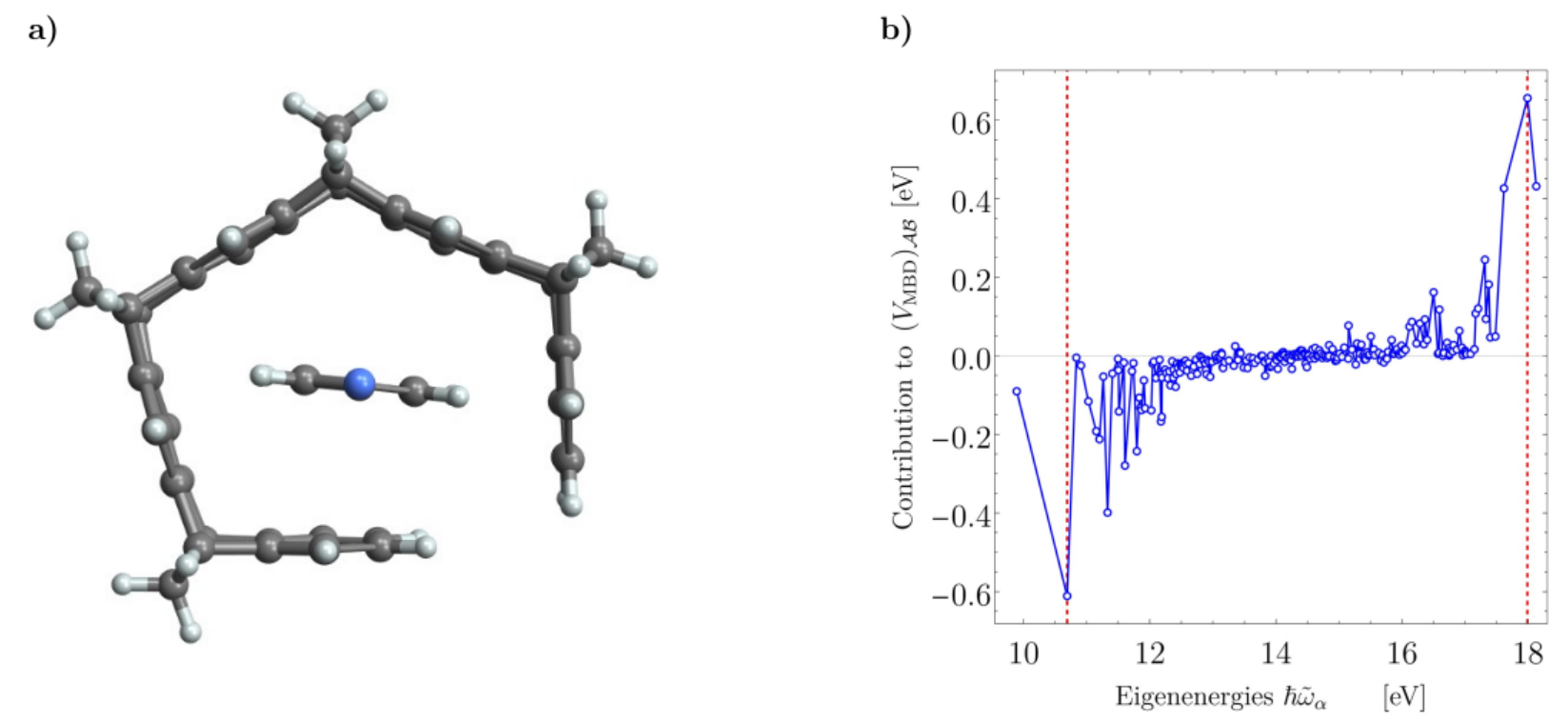}
    \caption{\textbf{Inter-fragment MBD potential energy $(V_{\rm MBD})_{\mathcal{A}\mathcal{B}}$ in the ``tweezer” complex with 1,4-dicyanobenzene, dominated by non-polar dispersion interactions.}.
    In panel a) the chemical structure of the complex \cite{grimme2012supramolecular} is shown, in panel b) the MBD eigenmode energies $\hbar \tilde{\omega}_{\rm MBD}$ are plotted vs. the contribution of each mode to the inter-fragment MBD potential energy $(V_{\rm MBD})_{\mathcal{A}\mathcal{B}}$. The red dashed lines highlight the MBD eigenenergies giving the most negative (left line) and most positive (right line) contributions to $(V_{\rm MBD})_{\mathcal{A}\mathcal{B}}$.}   
    \label{fig:S12L_2b_geoETmbd}
\end{figure}

In Figure~S\ref{fig:S12L_2b_geoETmbd} we have explored the contribution 
of each MBD mode to the interaction energy among fragments $(\hat{V}_{\rm MBD})_{\mathcal{A}\mathcal{B}}$ in the ``tweezer” complex dominated by 
non-polar dispersion interactions. Given the relatively small size of the complex ($\sim 10^2$ atoms), atomic resolution has been adopted. 
For this complex, the total MBD binding energy is $E_{\rm bind,MBD}\approx -0.818\,\,\mathrm{eV}$, while the inter-fragment interaction energy in the MBD ground state is $(V_{\rm MBD})_{\mathcal{A}\mathcal{B}}\approx -0.611 \,\,\mathrm{eV}$. This difference arises because in SQ-MBD we compute the expectation value of the inter-fragment interaction energy evaluated on the fully coupled MBD state. However, it is interesting to observe that these two alternative definitions of the binding energy are in relative agreement.
The right panel in Figure~S\ref{fig:S12L_2b_geoETmbd} shows the almost 
equal presence of positive contributions of ``anti-bonding" MBD modes and the negative contributions to the MBD interaction energy given by ``bonding" MBD modes. It is relevant to note the distribution in
the energy of these modes: the most negative one ($(V_{\rm MBD})_{\mathcal{A}\mathcal{B}}\approx -0.6 \,\,\mathrm{eV}$) given by a low-frequency MBD eigenmode ($\hbar\tilde{\omega}_{\rm min}\simeq 10.63 \,\,\mathrm{eV}$) and the most positive one ($(V_{\rm MBD})_{\mathcal{A}\mathcal{B}}\approx +0.6 \,\,\mathrm{eV}$) given by a high-frequency eigenmode ($\hbar\tilde{\omega}_{\rm max}\simeq 17.99 \,\,\mathrm{eV}$), while in the intermediate frequency range we observe a high density of modes giving an almost zero contribution to the interaction energy.

\begin{figure}[tbh!]
    \centering
    \includegraphics[scale=0.65]{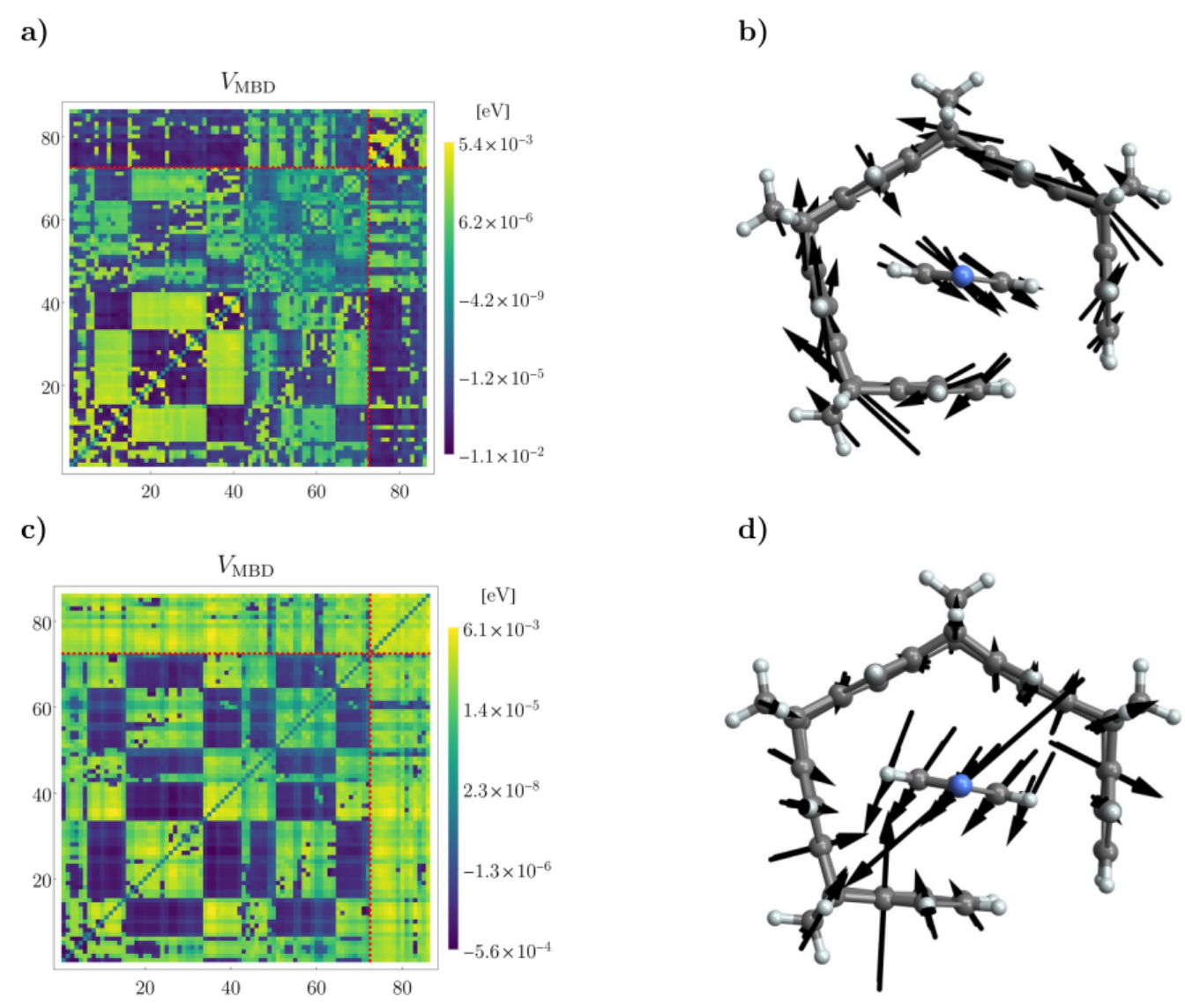}
    \caption{\textbf{Analysis of MBD eigenmodes contributions to $(V_{\rm MBD})_{\mathcal{A}\mathcal{B}}$ in the ``tweezer” complex with 1,4-dicyanobenzene, dominated by non-polar dispersion interactions.}
    In the upper panels, the $V_{\rm{MBD}}$ matrix at atomic resolution in a) and the components of the normalized MBD eigenvector in b) are reported for the MBD mode giving the
    most negative contribution to the inter-fragment MBD potential energy.
    In the lower panels, the $V_{\rm{MBD}}$ matrix at atomic resolution in c) and the components of the normalized MBD eigenvector in d) are reported for the MBD mode giving the most positive contribution to the inter-fragment MBD potential energy.
    The dashed (perpendicular) red lines in the $V_{\rm{MBD}}$ matrices separate the sets of atoms belonging to the two unlinked fragments in the complex.}   
    \label{fig:S12L_2b_modes}
\end{figure}

The normalized MBD eigenvectors for both the strongest ``bonding" and ``anti-
bonding" MBD modes are reported in panels on the right side of Figure~S\ref{fig:S12L_2b_modes}. The MBD eigenvector associated with the 
``bonding" MBD eigenmode $\tilde{\omega}_{\rm min}$ defines a 
``solenoidal"-like vector field over the atoms in the complex. 
In contrast, the MBD eigenvector associated with $\tilde{\omega}_{\rm max}$ seems to exhibit an ``irrotational"-like vector field. 
The SQ-MBD formalism allows one to easily estimate and visualize the 
contribution of each pair of atomic QDOs to $(V_{\rm MBD})_{\mathcal{A}\mathcal{B}}$ for a given MBD mode. 
In particular, it is possible to notice that in the eigenmode $\tilde{\omega}_{\rm min}$, the interaction energy between the two complexes is minimized, as demonstrated by the predominance of blue regions in Figure~S\ref{fig:S12L_2b_modes}a), but this is compensated by the increased MBD interaction energy among the atoms in the eigenmode $\tilde{\omega}_{\rm max}$, as demonstrated by the predominance of green and yellow regions in Figure~S\ref{fig:S12L_2b_modes}c).

\subsubsection{The electrostatically bound cucurbituril complex}

\begin{figure}[tbh!]
    \centering
    \includegraphics[scale=0.55]{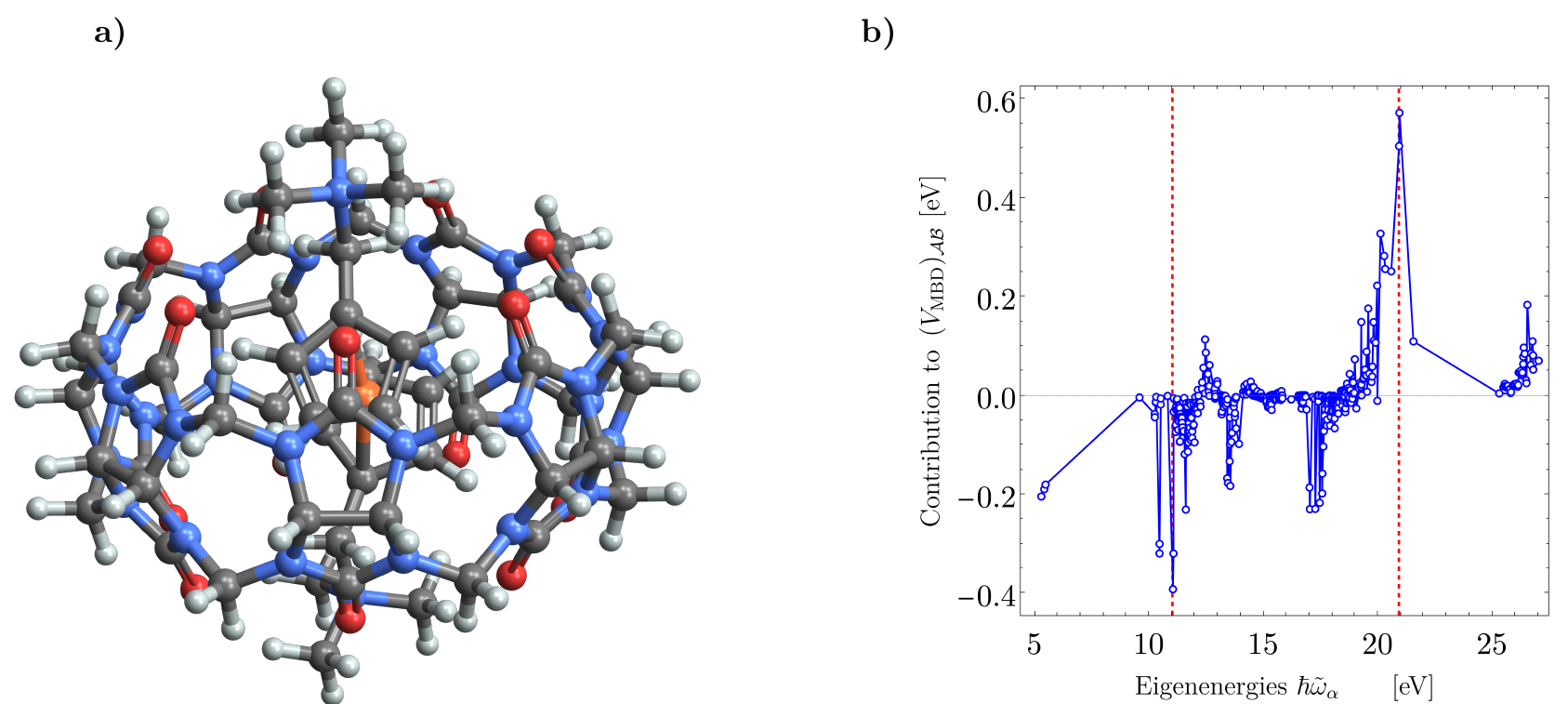}
    \caption{\textbf{Inter-fragment MBD potential energy $(V_{\rm MBD})_{\mathcal{A}\mathcal{B}}$ in the electrostatically bound cucurbituril complex}.
    In panel a) the chemical structure of the complex \cite{grimme2012supramolecular} is shown, with the iron ion $\rm{Fe}^{2+}$ displayed in orange at the center. In panel b) the MBD eigenmode energies $\hbar \tilde{\omega}_{\rm MBD}$ are plotted vs. the contribution of each mode to the inter-fragment MBD potential energy $(V_{\rm MBD})_{\mathcal{A}\mathcal{B}}$. The red dashed lines correspond to the MBD eigenenergies giving the most negative (left line) and most positive (right line) contributions to $(V_{\rm MBD})_{\mathcal{A}\mathcal{B}}$.} 
    \label{fig:S12L_7a_geoETmbd}
\end{figure}

\begin{figure}[tbh!]
    \centering
    \includegraphics[scale=0.1]{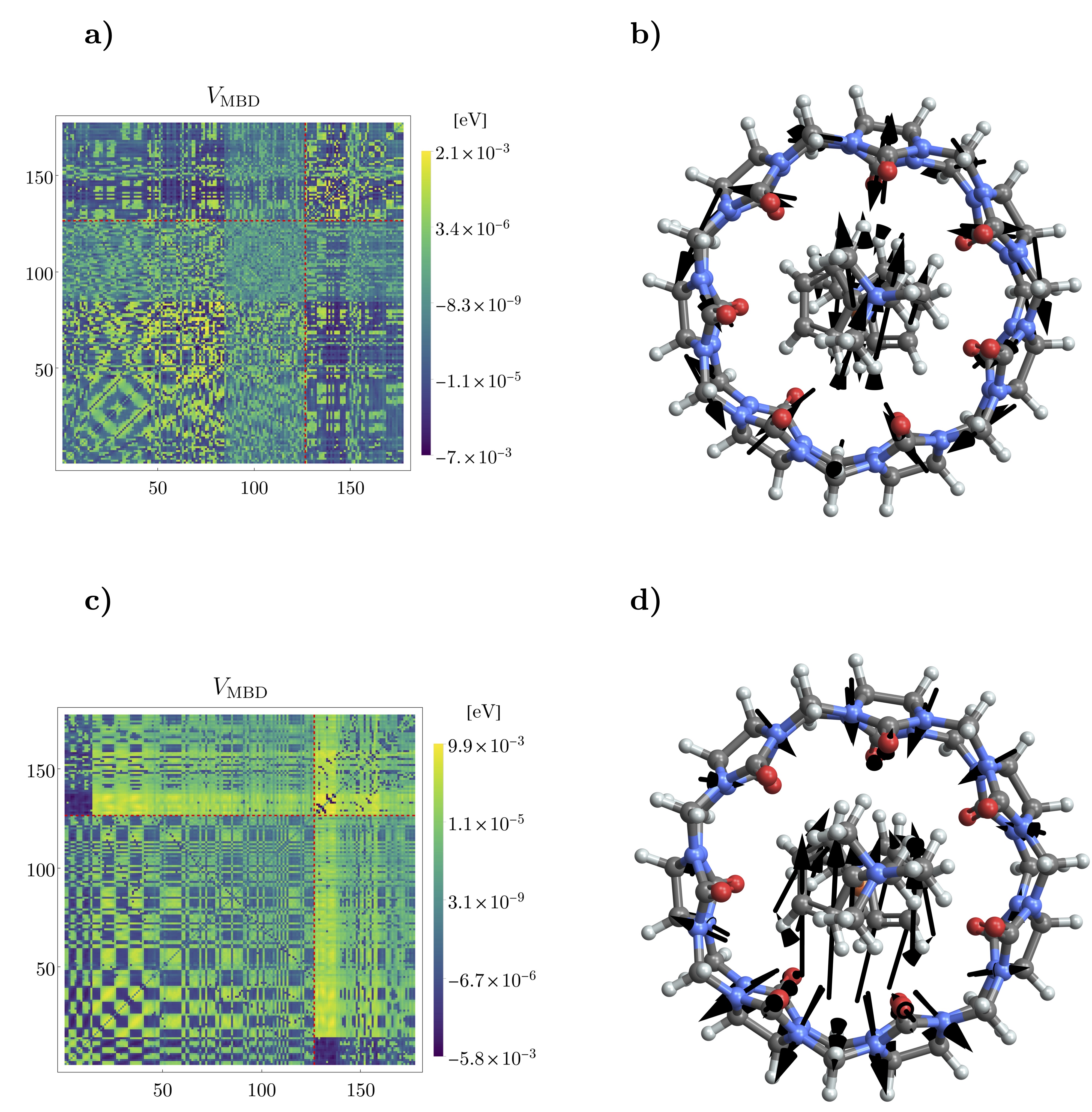}
    \caption{\textbf{Analysis of MBD modes contributions to $(V_{\rm MBD})_{\mathcal{A}\mathcal{B}}$ in the electrostatically bound cucurbituril complex.}
    In the upper panels, the $V_{\rm{MBD}}$ matrix at atomic resolution in a) and the components of the normalized MBD eigenvector in b) are reported for the MBD mode giving the
    most negative contribution to the inter-fragment MBD potential energy.
    In the lower panels, the $V_{\rm{MBD}}$ matrix at atomic resolution in a) and the components of the normalized MBD eigenvector in b) are reported for the MBD mode giving the
    most positive contribution to the inter-fragment MBD potential energy.
    The dashed (perpendicular) red lines in the $V_{\rm{MBD}}$ matrices separate the sets of atoms belonging to the two unlinked fragments in the complex.}  
    \label{fig:S12L_7a_modes}
\end{figure}

An analogous MBD modal analysis has been performed on 
the cucurbituril system, where electrostatic
interactions dominate the bound energy (see Figure~S\ref{fig:S12L_7a_geoETmbd}). In this case, the bonding energy between the cucurbituril ring and 
the smaller complex at its center  ($E_{\rm bind,MBD}\approx -1.890\,\,\mathrm{eV}$ )
and the MBD interaction energy ($(V_{\rm MBD})_{\mathcal{A}\mathcal{B}}\approx -1.865\,\,\mathrm{eV}$)
are almost equal.
Both ``bonding" and ``antibonding" MBD modes have been found,
with the most prominent ``bonding" mode ($(V_{\rm MBD})_{\mathcal{A}\mathcal{B}}\sim -0.4 \,\,\mathrm{eV}$) in 
the low-frequency region
at $\hbar\tilde{\omega}_{\rm min}\sim 11 \,\,\mathrm{eV}$ and the 
most prominent ``antibonding" mode ($(V_{\rm MBD})_{\mathcal{A}\mathcal{B}}\sim +0.6 \,\,\mathrm{eV}$) 
in the high-frequency region at $\hbar\tilde{\omega}_{\rm max}\sim 21 \,\,\mathrm{eV}$.
However, ``antibonding" MBD modes have also been
found in this case in a middle-low 
frequency range $\hbar\tilde{\omega}\approx 12.5 \,\,\rm{eV}$ associated with QDOs excitations among the terminus of the 
complex inside the ring, and ``bonding" MBD modes in the middle-high frequency range $\hbar\tilde{\omega}\approx 17.5 \,\,\rm{eV}$ associated with the collective excitation between each of the 
phenyl groups and half of the cucurbituril ring.
The SQ-MBD-based analysis of the most prominent ``bonding" MBD mode in Figure~S\ref{fig:S12L_7a_modes}a) 
shows a more homogeneous pattern of the matrix with respect to the 
previous ``tweezer" system. This can be
interpreted as a consequence of the high quasi-cylindrical 
symmetry of the system: the coherence between the atomic QDOs in 
the cucurbituril ring and the QDOs in the smaller complex does not prevent coherences
between the atomic QDOs in the fragments. Such an analysis is
complemented by the geometrical visualization provided
in Figure~S\ref{fig:S12L_7a_modes}b), where the normalized MBD 
eigenvector is reported on the real geometry, also showing in
this case a ``quasi-solenoidal" vector field.

On the contrary, the SQ-MBD-based analysis of the most prominent ``antibonding" MBD mode in Figure~S\ref{fig:S12L_7a_modes}c)  
reveals a strong coordination pattern among atoms in the larger complex that does not coordinate with the QDOs in the smaller complex at the center of the ring. Figure \ref{fig:S12L_7a_modes}d) reveals 
the geometry of the ``irrotational"-like vector
field of the normalized displacements associated with
the ``antibonding" modes.

\clearpage

\bibliographystyle{apsrev4-1}
\bibliography{apssamp.bib}